\documentclass[12pt]{spieman}  
\usepackage{amsmath,amsfonts,amssymb}
\usepackage{graphicx}
\usepackage{setspace}
\usepackage{tocloft}
\usepackage{lineno}
\usepackage{subcaption}
\usepackage{array}
\usepackage{wrapfig}
\usepackage{orcidlink}
\usepackage{xcolor, soul}
\usepackage{verbatim}
\usepackage{lineno}
\definecolor{mycolor}{RGB}{255,255,0} 
\sethlcolor{mycolor} 
\title{\underline{PyISH}: \underline{Py}thon \underline{I}ntegral \underline{F}ield Spectroscopy 
\underline{S}imulation for \underline{H}WO}

\author[a,b,c,d,e, *]{Grace Sweetak \orcidlink{0009-0000-6796-4056}}
\author[c]{Breann Sitarski}
\author[f]{Kevin France}
\author[e]{Randall McEntaffer}
\author[g]{Richard Cartwright}
\affil[a]{Lehigh University, 27 Memorial Dr W, Bethlehem, PA, United States}
\affil[b]{Southeastern Universities Research Association, Washington, DC, USA}
\affil[c]{NASA Goddard, 8800 Greenbelt Rd, Greenbelt, MD, United States}
\affil[d]{Center for Research and Exploration in Space Science and Technology, NASA/GSFC, Greenbelt, MD, USA}
\affil[e]{Pennsylvania State University, 201 Old Main, University Park, PA, United States}
\affil[f]{Laboratory for Atmospheric and Space Physics, University of Colorado Boulder, Boulder, CO 80309}
\affil[g]{Johns Hopkins University Applied Physics Laboratory, Laurel, Maryland, USA}

\cftpagenumbersoff{figure}
\cftpagenumbersoff{table} 
\begin{document} 
\maketitle

\begin{abstract}

The Habitable Worlds Observatory (HWO) will be a large ultraviolet/optical/near-infrared space telescope operating at the Sun-Earth Lagrange point L2. HWO was highly recommended by the National Academies’ 2020 decadal survey and will be the first telescope designed specifically to search for life on planets orbiting other stars. HWO will also be able to perform a host of other transformational astrophysics, including cosmology, galaxy evolution, solar system science, and beyond. The development of the telescope and instrument suite is an iterative process. Example observatory architectures, called exploratory analytic cases (EACs) by the HWO Technology Maturation Project Office (TMPO), are modeled end-to-end to explore the engineering and science trade space. Recently, an ultraviolet Integral Field Spectrograph (UV IFS) was added to HWO's instrument suite for the EACs 4 and 5. To explore the science and engineering trade space for this specific instrument, we developed a high-fidelity UV IFS simulation tool, PyISH. The UV IFS simulation tool is designed to be used by scientists to model specific science cases as seen by a UV IFS on HWO, as well as engineers to explore the trade space when developing potential instrument architectures. The modular components, deliverables, and an example of the tool simulating a specific science case proposed for HWO are described in this paper.

\end{abstract}


{\noindent \footnotesize\textbf{*}Grace Sweetak,  \linkable{grace.m.sweetak@nasa.gov} }

\begin{spacing}{2}   

\section{Introduction}
\label{sect:intro}

\begin{figure}
    \centering
    \includegraphics[width=0.8\linewidth]{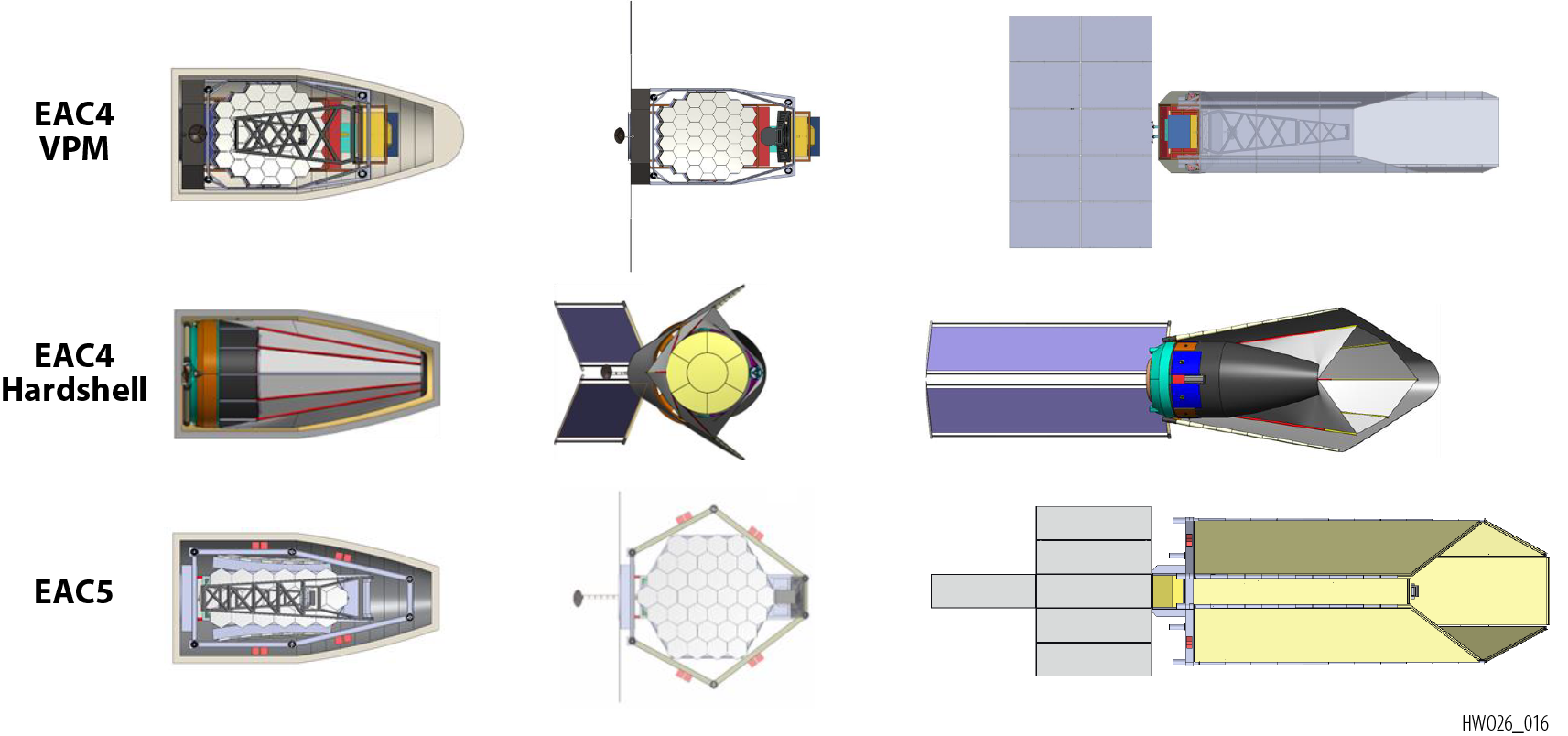}
    \caption{Exploratory Analytic Cases (EACs) 4 and 5. Both designs will include a UV IFS, with EAC 4 having an inscribed primary diameter greater than 6.5 meters, and EAC 5 having an inscribed primary diameter greater than 8 meters {\cite{Feinberg2026HabitableExploration}}.}
    \label{figs: eacs}
\end{figure}

HWO is the top recommendation for large astrophysics missions by the National Academies’ Decadal Survey on Astronomy and Astrophysics 2020 (Astro2020) to search for signs of life orbiting other stars, and undertake many other transformational astrophysics science cases{\cite{NationalAcademies2023Pathways2020s}}. Several exploratory analytic cases (EACs), or preliminary architecture concepts for HWO, are being studied to understand key architectural options and breakpoints, practice end-to-end modeling, and identify key technology gaps and solutions        {\cite{Feinberg2024TheOpportunities}}. Figure \ref{figs: eacs} depicts the preliminary concepts for the most recent EACs 4 and 5, the newest architectures that the HWO Technology Maturation Project Office (HWO TMPO) is studying.

The two telescopes vary in size, with EACs 4 and 5 having an inscribed primary diameter of $\geq$ 6.5 and 8 meters, respectively. Both EACs contain 5 instruments. The UV IFS, which was not included in EACs 1 - 3, was recently added to EACs 4 and 5 based on community interest from a wide range of science case development documents (SCDDs) submitted by the community to HWO TMPO (Dressing et al. (in prep)). In fact, $\approx$ 25 \% of the SCDDs expressed interest in using a UV IFS to perform their science. The 25\% of these SCDDs cover a diversity of science cases, spanning from solar system science, CGM emission, and supernova remnants. The UV IFS’s architecture will be explored as part of EACs 4 and 5 in the coming months. For more information on EACs 4 and 5, please see Feinberg et al 2026 {\cite{Feinberg2026HabitableExploration}}. 

An IFS provides both spatial and spectral information to the user in the form of a three-dimensional data cube. The submitted SCDDs call for a wide range of performance parameters for the UV IFS, including a field of view between 1 - 10 arcseconds, a spectral resolving power between 100 - 10,000, and the angular size of each spaxel (three dimensional spatial pixel) between 10-50 milliarcseconds {\cite{Scowen2025TechnologyObservatory}}.

NIRSpec and MIRI both have IFS modes on JWST, but the only UV IFS flown in space as of 2026 was the Integral Field Ultraviolet Spectroscopic
Experiment (INFUSE), a sounding rocket launched in 2023 that will fly again in 2026 and 2027 {\cite{Boker2022TheTelescope,Haughton2025INFUSESpectrograph}}. Due to the wide range of IFS performance desires from various science cases, and the early stages of UV IFS technology maturation, we developed a UV IFS simulation to explore the science and engineering trades that could be accommodated for current and future EACs for HWO. 

Previous observatories that include an IFS created simulations to complement and/or prepare for the instrument. \texttt{Specsim} was developed in preparation for the IFS mode on the JWST MIRI instrument, and the instrument performance simulator was developed for the NIRSpec instrument and includes an IFS mode{\cite{Lorente2006Specsim:Simulator, Piqueras2008TheSimulator}. The Roman Space Telescope Coronagraph Instrument’s de-scoped IFS mode was simulated with the \texttt{crispy} software package{\cite{Rizzo2017SimulatingSpectrograph}}. More recently, an IFS simulation was developed for the Chinese Space Station Survey Telescope (CSST), scheduled to launch in 2027{\cite{Yan2026MockSimulation}}. IFS simulations accommodate ground-based observatories as well, with \texttt{HSIM} as the complementary simulation tool for the HARMONI integral field spectrograph on the European Extremely Large Telescope, for example{\cite{Zieleniewski2015HSIM:ELT}}. Comparing all of the IFS simulations, PyISH is the only UV simulation and previous simulations differ instrument specifications or overarching design.
Our simulation, allows users to experiment with different IFS specifications such as detector size, plate scale, spectral resolving power, etc. \texttt{Crispy} allows for limited modularity, however, the overarching design was a lenslet-based IFS, rather than PyISH's slicer-based IFS design, which requires a different simulation approach. Our slicer-based design choice for PyISH is justified below. }

In addition to the instrument specification trades with PyISH, our simulation allows members of the scientific community to understand the power of a UV IFS by simulating their specific science case as observed through a UV IFS on HWO. Therefore, the HWO UV IFS simulation serves two purposes: 

\begin{enumerate}
    \item Explore the science and engineering trades for including a UV IFS in the HWO instrument suite.
    \item Enable the community to visualize UV IFS data products.
\end{enumerate}

\newpage

\begin{figure}[h!]
    \centering
    \includegraphics[width=0.8\linewidth]{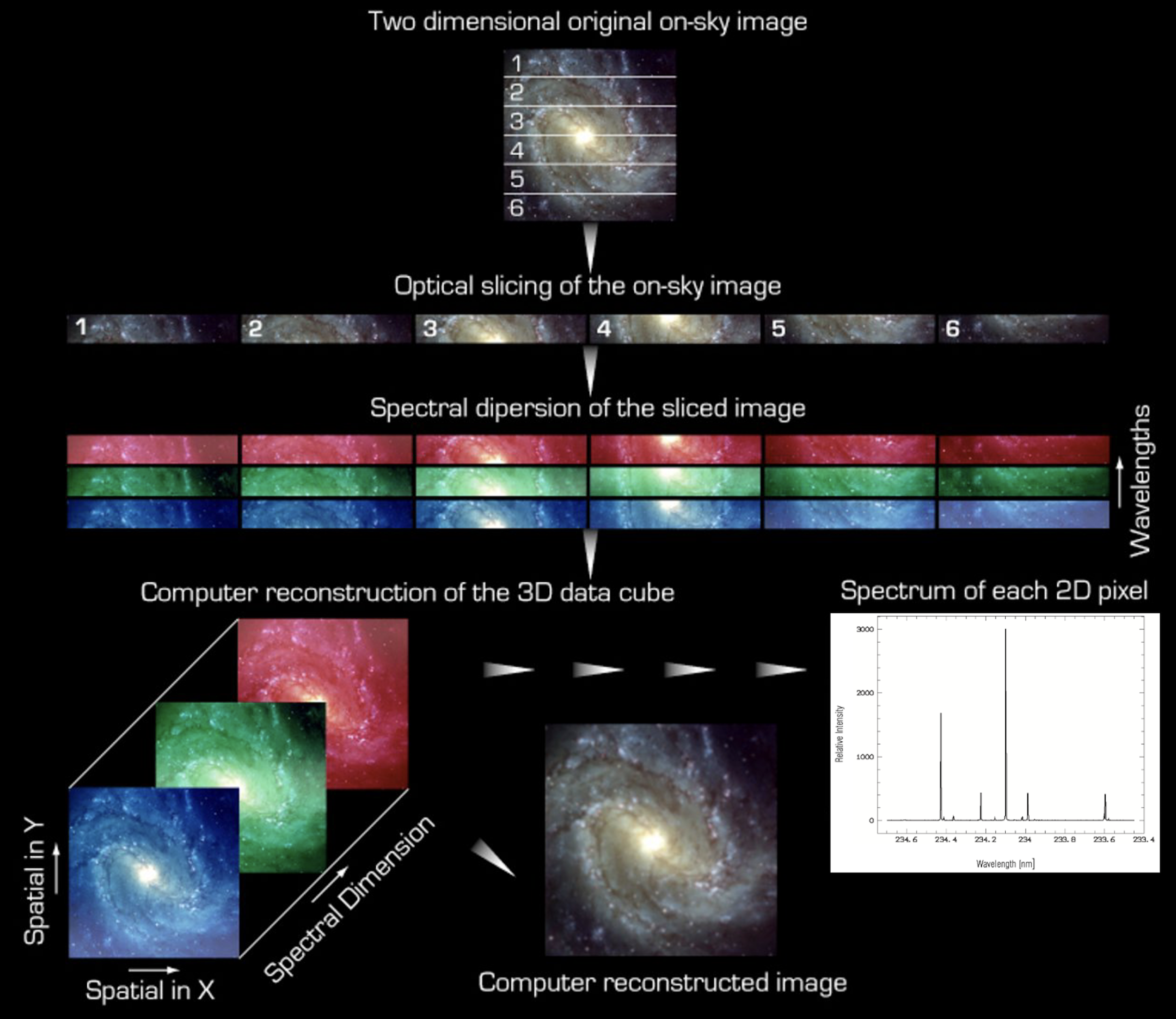}
    \caption{Schematic depicting the function of each optic in an Integral Field Spectrograph  (IFS){\cite{EuropeanSouthernObservatoryTechnologyUnits, SpaceTelescope:EuropeanCoordinatingFacility2005ST-ECFNewsletter}}. }
    \label{fig: description IFS}
\end{figure}

A slicer-based IFS is an optimal choice for UV to ensure high throughput. Alternative IFS designs such as lenslet-based designs or fiber bundles do not historically have high transmission in the UV, although development is ongoing{\cite{Tuttle2024UltravioletObservatory}}. Low UV transmission, coupled with the success of a slicer-based IFS with INFUSE led to our decision to simulate a reflective slicer-based UV IFS over other design options. 

A general overview of a slicer-based IFS is shown in the schematic in Figure \ref{fig: description IFS}. Light from an astronomical source passes through the telescope to the UV IFS, where it is spatially sliced using a reflective slicer element.  Each slice of the image is then sent to its own grating in a grating array. In Figure {\ref{fig: description IFS}}, the grating array consists of 6 gratings. The dispersed slices are captured on different sections of the detector. Post-processing techniques are used to reconstruct the 3D data cube, which provides a spectrum of each spaxel. 

The following sections will describe the implementation of each component of the UV IFS to the simulation tool. Section \ref{sect: simulation tool modul es} describes each module in the simulation tool, including the assumptions and user inputs. Section \ref{sect: europa example} shows an example of the simulation tool modeling the science case, ``Do large icy bodies and moons harbor habitable environments in their interiors?" Section \ref{sect: discussion/future work} describes how scientists and engineers can utilize our tool for a variety of cases in the context of HWO.

\section{Simulation Tool Modules}
\label{sect: simulation tool modul es}
\begin{figure}[h!]
    \centering
    \includegraphics[width=0.95\linewidth]{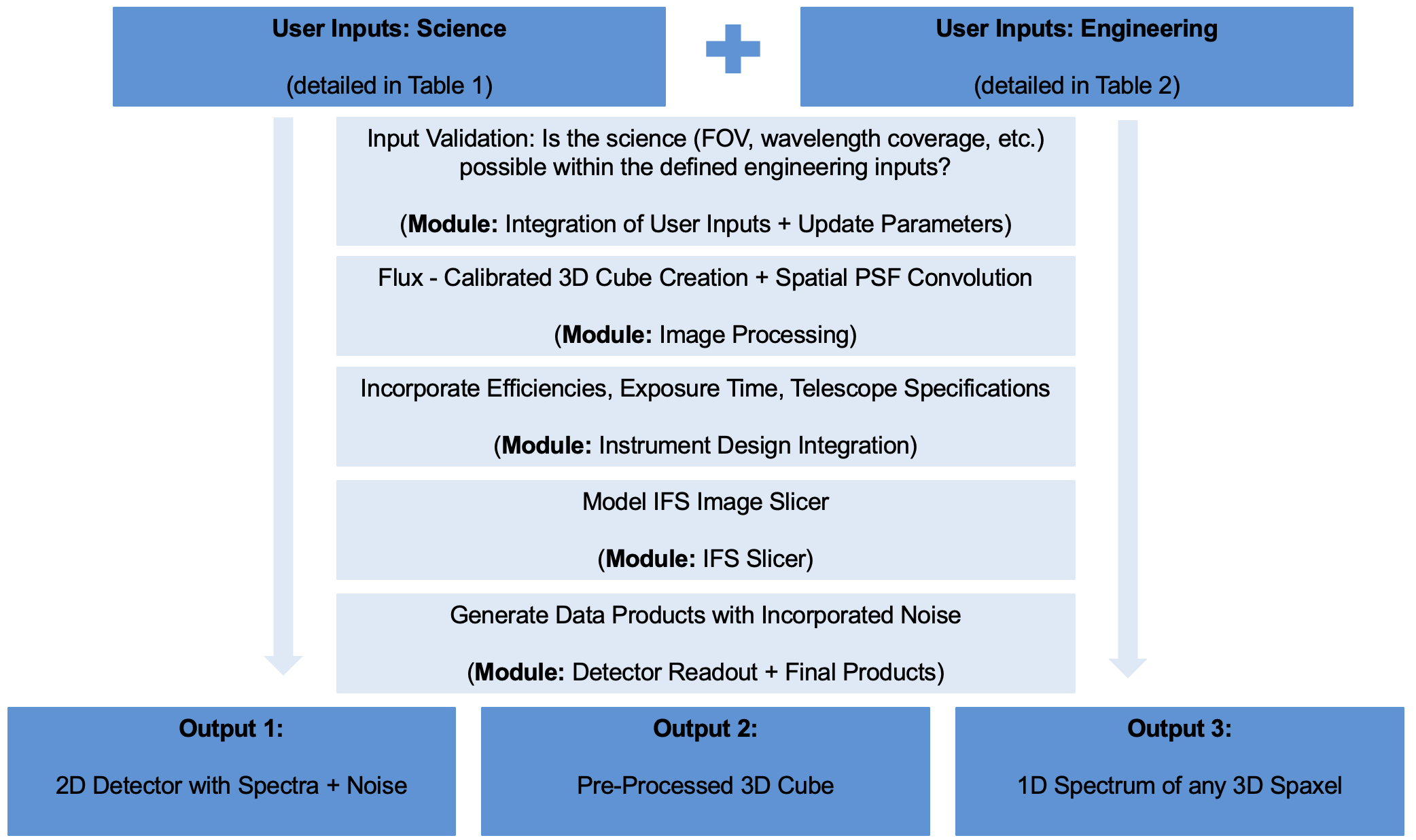}
    \caption{Flowchart describing the goal of each module in our UV IFS simulation.}
    \label{fig: Flowchart}
\end{figure}

 Each subsection within this section describes a different module in the simulation. Figure \ref{fig: Flowchart} depicts a flowchart through the simulation and through each module.

\subsection{Integration of User Inputs}
\label{subsect: integration user inputs}

The user-provided inputs needed to run the simulation are shown in Tables \ref{table: user inputs, science} and \ref{table: user inputs, engineering}. The first table describes the science-related inputs, and the second table describes the engineering inputs of the desired instrument architecture.

\begin{table}[h!]
\begin{center}
\begin{tabular}{||c c||} 
 \hline
  Science Input & Description \\ [0.5ex] 
 \hline\hline
  Name & List of names of each emission line \\ 
  \hline
 $\lambda_{emission}$ & List of wavelength centroid of each ion [nm]  \\
 \hline
 $\lambda_{width}$ & List of widths of each emission line peak in given spectrum [nm] \\
 \hline
 Spatial maps & List of high resolution spatial maps for each given emission line [.fits] \\
 \hline
 Spectrum & Spectrum of source [.txt file, Rayleighs vs. nm]\\
 \hline
 Continuum Map & Spatial map of continuum  [.fits]\\
 \hline
 $FOV_{spectra}$ & One dimension of spatial maps (assume square) [arcsec] \\ [1ex] 
 \hline
\end{tabular}
\end{center}
\caption{Science specific inputs that the user specifies for the UV IFS simulation. The maps are given as paths to specific files. For an example of each science input, see Section \ref{subsec: integration europa}.  }
\label{table: user inputs, science}
\end{table}

\begin{table}[h!]
\begin{center}
\begin{tabular}{||c c||} 
 \hline
  Engineering Input & Description \\ [0.5ex] 
 \hline\hline
  $N_{slices}$ & Number of IFS slices \\ 
  \hline
 $dim_{detector, 1}$ & One dimension of the detector [mm] \\
 \hline
 $dim_{detector, 2}$ & One dimension of the detector [mm] \\
 \hline
 $\omega_{dark}$ & Dark rate of detector [cts/cm$^2$/s] \\
 \hline
 $\lambda$ & Wavelength used to calculate $\Delta\lambda$ (see Equation \ref{equ: delta lambda}) [nm] \\
 \hline
 $R_{\lambda}$ & Spectral resolving power for the given $\lambda$ [unitless] \\
 \hline
 $bp$ & Minimum bandpass desired [nm]\\
 \hline
 $FOV_{IFS}$ & One dimension of the minimum field of view desired (assume square) [arcsec] \\
 \hline
 $\Omega_{pix}$ & Plate scale [mas/pixel] \\
 \hline
 $\gamma$ & Pixel pitch [$\mu m$/pixel] \\
 \hline
$D_{ins}$ & Inscribed diameter of primary mirror [meters] \\
 \hline
 $T_{exp}$ & Exposure time [sec] \\
 \hline
 $\lambda_{start}$ & Low cutoff bandpass [nm] \\ 
 \hline
 $N_{ref}$ & Number of reflections in telescope and IFS design [unitless] \\  
 \hline
 $EAC_{paths}$ & path to each EAC file [.yaml]  \\ 
 \hline
 $EAC$ & EAC to simulate  \\ [1ex] 
 \hline
\end{tabular}

\end{center}
\caption{Engineering specific inputs that the user specifies. For an example of each engineering input, see Section \ref{subsec: integration europa}.}
\label{table: user inputs, engineering}
\end{table}

The science inputs pertain to a specific science case that the user wants to simulate and include specifying the wavelength range for spectral features, spatial maps of those features, a spatial representation of the continuum, and a 1D spectrum for flux calibration. The spatial maps are high resolution images or models corresponding to a specific wavelength listed in $\lambda_{emission}$, while the continuum map is a spatial image or model of the source at wavelengths corresponding to the continuum.
Discussion on the implementation of these inputs begins in Section \ref{subsect: Image Processing}.

The engineering inputs take into account the physical constraints of the HWO UV IFS, many of which are based on the size and characteristics of the detector. A larger detector means more available space for spectra, meaning more or longer IFS slices, or a longer bandpass. The current simulation gives the user an option whether to use the entirety of the detector input size that is given in Table \ref{table: user inputs, engineering}, or to determine and use the minimum detector size needed to achieve the $R$, bandpass, and $FOV_{IFS}$ constraints in Table \ref{table: user inputs, engineering}. 

Using the entire detector size means that the user could have a much larger field of view and/or bandpass than required for the specific science case. This would occur if the user's requested detector size is much bigger than what is needed to achieve the spectral resolving power, bandpass, and field of view constraints in Table \ref{table: user inputs, engineering}. It is useful to understand the tradeoff between the detector size needs for various science cases and what is available as the current state-of-the-art. With this optimization tool, the tradeoffs between the desired constraints of various science cases, and different optical designs can be explored.

However, using all of the user-specified detector area can be suboptimal, for example, a larger field of view would increase the probability of bright background stars saturating the detector in longer exposures. Consequently, to explore engineering vs. science trades, the user has the option to not use all of the detector space. In this case, the simulation will change the detector size to be the minimum size needed to achieve the requested spectral resolving power, bandpass, and IFS field of view. Changing to the ``minimum detector size needed" is equivalent to only using a portion of a larger detector. The choice between using a portion of the detector and using the full space given by the user is discussed further in Section \ref{subsect: Update parameters}.

Regardless of whether a user chooses to use all of the detector space, any detector on an IFS will be sectioned into different regions pertaining to different slice-grating pairs. As seen in Figure \ref{fig: description IFS}, each slice of the on-sky image is dispersed with a different grating. Each spectrum needs to be captured on a different region of the detector to avoid overlapping with neighboring spectra.

To partition the detector into the needed regions, the simulation converts the detector size given in Table \ref{table: user inputs, engineering} to pixels using Equation \ref{equ: detector in pixels}, 

\begin{equation}
    N_{pix, detector} = \frac{dim_{detector}}{\gamma},
    \label{equ: detector in pixels}
\end{equation}

where both $dim_{detector}$ and $\gamma$ are given in Table \ref{table: user inputs, engineering}. Once the detector size in pixels is determined, the minimum sizes for each section are calculated. The minimum width of each section of the detector is calculated using Equation \ref{equ: min x in pixels}, 

\begin{equation}
    x_{min, pixels} = \frac{bp}{ \Delta\lambda} * 2,
    \label{equ: min x in pixels}
\end{equation}

where $\Delta\lambda$ is

\begin{equation}
    \Delta\lambda = \frac{\lambda}{R_{\lambda}}.
    \label{equ: delta lambda}
\end{equation}

The factor of 2 in Equation \ref{equ: min x in pixels} accounts for Nyquist sampling. The width of each section is determined by the spectral constraints; the longer the bandpass, or the greater the resolution, the larger space needed on the x-dimension of the detector. The minimum height is associated with the spatial dimension. It is determined using Equation \ref{equ: min y in pixels}, 

\begin{equation}
    y_{min, pixels} = \frac{FOV_{IFS}}{\Omega_{pix}},
    \label{equ: min y in pixels}
\end{equation}

where both $FOV_{IFS}$ and $\Omega_{pix}$ are given by the user in Table \ref{table: user inputs, engineering}. It is important to note that in both Table \ref{table: user inputs, engineering} and Equation \ref{equ: detector in pixels}, the detector axes are not assigned an x or y orientation. They are assigned an orientation based on Equations \ref{equ: min x in pixels} and \ref{equ: min y in pixels}; whichever equation results in the higher value will be assigned the larger detector dimension, in the case that the detector is rectangular. Equation \ref{equ: min x in pixels} depends on bandpass and $\Delta\lambda$, as the x-direction specifies the spectral dimension, while Equation \ref{equ: min y in pixels} depends on $FOV_{IFS}$, specifying the spatial direction. Thus, we make the choice in assigning an x and y orientation to give more space to the dimension that requires it to meet science needs. 

 After determining the minimum sizes of each section needed to fulfill the user's constraints and the orientation of the detector, sections are initially divided through the detector by making even cuts on the x and y axis. For example, if the user specified 81 IFS slices, there would be 9 sections across the x axis, and 9 sections across the y axis. We place an additional check to ensure that the amount of sections totals or exceeds the number of IFS slices specified by the user. Each IFS slice results in a distinct spectrum that must be placed in a separate section of the detector, requiring at a minimum the same amount of sections as requested IFS slices. As an example, if the user specified 82 IFS slices instead of 81, the simulation would start with a 9x9 section setup, and then add an additional column to result in 10x9 sections, exceeding the amount of slices requested. 

Next, the simulation determines the minimum overall detector size needed given the orientation of sections. The minimum detector size is determined using Equations \ref{equ: min x detector} and \ref{equ: min y detector}, 

 \begin{equation}
     N_{x_{pix},detector_{min}} = x_{min, pixels} *N_{sections,x}
     \label{equ: min x detector}
 \end{equation}

 \begin{equation}
     N_{y_{pix},detector_{min}} = y_{min, pixels} *N_{sections,y}
      \label{equ: min y detector}
 \end{equation}

 where $x_{min, pixels}$ and $y_{min, pixels}$ come from Equations \ref{equ: min x in pixels} and \ref{equ: min y in pixels}, respectively, and $N_{sections,x}$ and $N_{sections,y}$ are the number of sections across the x and y axes, respectively. 

 At this point, we developed different logical paths the simulation can follow depending on the result of Equations \ref{equ: min x detector} and \ref{equ: min y detector}. If both of these values are smaller than the user-defined dimensions for the detector that the user input given in pixels by Equation \ref{equ: detector in pixels}, the given detector will meet the minimum science constraints and the simulation will continue to the logic described in Section \ref{subsect: Update parameters}.

 If one or both of the results from  Equations \ref{equ: min x detector} and \ref{equ: min y detector} are larger than the given detector size, the current orientations of the sections in the detector do not meet the science constraints. The simulation automatically adjusts the orientation of the sections in the allotted detector space to attempt to reshape the individual sections. For example, if the sections for each spectra are too narrow, we attempt to reshape how the detector sections are organized to make each section shorter and wider. This logic is described in Options A, B, and C in the following sections.

\subsubsection{Option A: Adjust Height}
\label{subsubsect: option A: Adjust Height}

Option A is implemented when the given detector width, or $N_{pix, detector}$ for the x-axis, is larger than the minimum detector width needed as determined by Equation \ref{equ: min x detector}, but the given detector height, or $N_{pix, detector}$ for the y-axis, is smaller than the minimum detector height needed. This means that the bandpass and desired spectral resolution will fit on the width allocated by sectioning the detector as described in Section \ref{subsect: integration user inputs}, but the field-of-view (FOV) desired will not fit on the space allocated in the y-axis. Therefore, Option A attempts to allocate more space on the detector's y-axis. 

Allocating more space on the y-axis requires an iterative process of removing rows and adding columns: in every iteration, one row is removed and one column is added. Then, a check is implemented to ensure the total sections meets or exceeds the amount of IFS slices, as done in Section \ref{subsect: integration user inputs}. In the case that the total sections no longer meet or exceed the number of IFS slices, additional columns are added until the requirement is met. After adjusting the number of rows and columns, the minimum detector size needed is recalculated, again by using Equations \ref{equ: min x detector} and \ref{equ: min y detector}. For each equation, the minimum section size stays the same, but the number of sections across the x-axis and the number of sections across the y-axis changes based on the new orientation. Finally, these updated minimum detector dimensions are checked against the given detector size from Table \ref{table: user inputs, engineering} that has been converted to pixels using Equation \ref{equ: detector in pixels}. This process is continued iteratively until the minimum detector dimensions in both x and y are smaller than the given detector dimensions. 

In summary, this process trades space on the x-axis for space on the y-axis, making each section narrower and taller. In the iterative process, there is the possibility that by allocating more space on the y-axis for the spatial constraints, the detector no longer has enough space on the x-axis to meet the spectral constraints. Every iteration has an additional check for the space remaining on the x-axis. If an iteration no longer allows the x-axis of each section to have enough space to meet the spectral constraints, the simulation stops, printing out the following statement, \texttt{"Changing the orientation of the cutouts to achieve spatial constraints voided the spectral constraints needed. The only option to achieve both is to use a larger detector."}

In the other case, where reorienting the sections within the detector results in the desired outcome that both the spectral and spatial constraints are met, the simulation will print, \texttt{"After adjusting spacing, this detector size works for your given constraints!"} and move on to Section \ref{subsect: Update parameters}.

\subsubsection{Option B: Adjust Width}
\label{subsubsect: option B: Adjust Width}

Option B is implemented by the detector when the given detector height, or $N_{pix, detector}$ for the y-axis is larger than the minimum detector height needed as determined by Equation \ref{equ: min y detector}, but the given detector width, or $N_{pix, detector}$ for the x-axis, is smaller than the minimum detector width needed. In this case, the FOV will fit into the space allocated on the y-axis, but the bandpass and spectral resolution desired will not fit on the width allocated by sectioning the detector as described in Section \ref{subsect: integration user inputs}. Therefore, Option B attempts to allocate more space on the detector's x-axis.

As in Option A described in Section \ref{subsubsect: option A: Adjust Height}, allocating more space on the x-axis is done by removing a column, and adding a row. Within each individual section, this reduces space on the y-axis and adds space on the x-axis, resulting in shorter, wider sections. As in Option A, the same process is followed in which sections are checked after reorganizing to ensure they meet or exceed the number of IFS slices, i.e. the number of spectra needed to fit on the detector. Then, the minimum detector size is recalculated using Equations \ref{equ: min x detector} and \ref{equ: min y detector}, and checked against the given detector size from Table \ref{table: user inputs, engineering}. This process is continued iteratively until the minimum detector dimensions in both x and y are smaller than the given detector dimensions. 

If by attempting to allocate space in the x-axis results in voiding the minimum space needed to achieve the science constraints on the y-axis, the simulation stops and a similar statement to Option A is printed, \texttt{"Changing the orientation of the cutouts to achieve spectral constraints voided the spatial constraints needed. The only option to achieve both is to use a larger detector."}

Otherwise, if reshaping the sections to shorter, wider sections to achieve both the spectral and spatial constraints is successful, the simulation will print, \texttt{"After adjusting spacing, this detector size works for your given constraints!"} and move on to Section \ref{subsect: Update parameters}.

\subsubsection{Option C: Adjust Height and Width}
\label{subsubsect: Option C: Adjust Height and Width}

Option C is invoked when both the given detector dimensions for x and y are smaller than the minimum detector dimensions calculated in Equations \ref{equ: min x detector} and \ref{equ: min y detector}. In this case, no trades can be made between the x and y axis, as allocating space for the spectral dimension will always reduce the needed space for the spatial dimension, and vice versa. Therefore, this option will print out the statement, \texttt{"Given the user inputs, I can not section the detector appropriately to fit all of the desired spectra. So, I cannot continue. Either change the science constraints or provide a larger detector and rerun."}

\subsubsection{Option Summary}
\label{subsubsect: Option Summary}

Table \ref{table: option summary} provides a summary of the logic Section \ref{subsect: integration user inputs} follows depending on the user-specified inputs.

\newpage

\begin{table}[h!]
\begin{center}
\begin{tabular}{ | m{5cm} | m{5cm}| m{5cm} | } 
  \hline
  Option & Option Logic & Possible Outcome \\ [0.5ex]
  \hline
  \hline
  Adequate space on detector's x or y axes & No further steps & Successful: move to Section \ref{subsect: Update parameters} \\ 
  \hline
 Option A:  Not enough space on detector's y-axis & Attempts to allocate space by removing rows and adding columns & Successful: move to Section \ref{subsect: Update parameters}. Unsuccessful: need larger detector \\
 \hline
Option B: Not enough space on detector's x-axis  & Attempts to allocate space by removing columns and adding rows & Successful: move to Section \ref{subsect: Update parameters}. Unsuccessful: need larger detector\\ 
\hline
Not enough space on detector's x or y axes & No further steps & Unsuccessful: need larger detector \\ 
  
 \hline
\end{tabular}

\end{center}
\caption{Summary of Sections \ref{subsect: Update parameters}, \ref{subsubsect: option A: Adjust Height}, \ref{subsubsect: option B: Adjust Width}, and \ref{subsubsect: Option C: Adjust Height and Width} describing the logic to section the space on the detector to achieve the user-defined science constraints.}
\label{table: option summary}
\end{table}

The result of ``Integration of User Inputs" module described in Section \ref{subsect: integration user inputs} is an appropriately sectioned detector according to the user specified engineering inputs for the UV IFS. 

\subsection{Update Parameters}
\label{subsect: Update parameters}

Once an orientation for the detector is determined using the processes described in Section \ref{subsect: integration user inputs}, the user chooses whether to use the entire space available on the detector. If the user chooses to keep the same detector size, the detector is sectioned in the orientation described in Section \ref{subsect: integration user inputs}. A check is done to ensure every section is the exact same dimensions, and cut a column or row of pixels off of particular sections if not. An updated FOV is calculated using Equation \ref{equ: hwo FOV}, 

\begin{equation}
    FOV_{HWO, full} =  \Omega_{pix} * y_{section, pix},
    \label{equ: hwo FOV}
\end{equation}

where $\Omega_{pix}$ is given in Table \ref{table: user inputs, engineering}, and $y_{section, pix}$ is the actual height in pixels of each section on the detector. An updated bandpass is determined using Equation \ref{equ: bandpass hwo},

\begin{equation}
    bp_{IFS} = \frac{\Delta\lambda}{x_{section,pix}} * \frac{1}{2}
    \label{equ: bandpass hwo}
\end{equation}

where $\Delta\lambda$ is determined in Equation \ref{equ: delta lambda}, and $x_{section, pix}$ is the actual width in pixels of each section of the detector. The factor of 1/2 accounts for Nyquist sampling.

In addition to updated science constraints, the minimum detector size needed to achieve the user input science constraints from Table \ref{table: user inputs, engineering} is calculated using Equations \ref{equ: min detector x update} and \ref{equ: min detector y update}, 

\begin{equation}
    x_{pix, min} =  N_{detector, x_{pix}} *\gamma
    \label{equ: min detector x update}
\end{equation}

\begin{equation}
    y_{pix, min} =  N_{detector, y_{pix}} *\gamma
    \label{equ: min detector y update}
\end{equation}

where $N_{x_{pix},detector_{min}}$ and $N_{y_{pix},detector_{min}}$ are from Equations \ref{equ: min x detector} and \ref{equ: min y detector}, respectively. These values were potentially updated if the simulation invokes resizing options A or B, as described in Sections \ref{subsubsect: option A: Adjust Height} and \ref{subsubsect: option B: Adjust Width}, respectively.

If the user chooses to keep the same science constraints and change the detector size, the results of Equations \ref{equ: min detector x update} and \ref{equ: min detector y update} are used as the new detector size, rounded up to the nearest integer. The same updated constraints for FOV and bandpass are printed using Equations \ref{equ: hwo FOV} and \ref{equ: bandpass hwo}. With the exception of minor rounding errors, the FOV and bandpass in this case are the same as the user-provided inputs. 

The final result of the ``Update Parameters" module is a sectioned detector, either fitting the maximum science return for FOV, bandpass, and resolution onto the given detector size, or fitting the minimum user-specified FOV, bandpass, and resolution specified in Table \ref{table: user inputs, engineering} onto the smallest detector size possible.

\subsection{Image Processing}
\label{subsect: Image Processing}

Throughout the simulation, we use inputs described in Table \ref{table: user inputs, science} to model any science case that would be detected by a UV IFS on HWO. In this specific module, we use the spectrum, spatial maps, and continuum maps the user provides to construct a flux calibrated 3D data cube across the user-specified bandpass. 

Prior to flux calibration, we convolve the normalized spatial maps with a point spread function (PSF) generated via the Physical Optics Propagation in PYthon (POPPY) simulation at each wavelength of interest from a set of telescope parameters{\cite{Perrin2012SimulatingWebbPSF}}. The telescope parameters are specified by the EAC. In Table \ref{table: user inputs, engineering}, the user specifies the location of the EAC parameter .yaml file in their home directory, as downloaded from the HWO Science and Engineering Interface Github. A PSF is generated at the FOV and plate scale specified by the user in Table \ref{table: user inputs, engineering} at each wavelength and normalized to the peak flux in the PSF. To date, an IFS-specific line spread function (LSF) has not been created for EACs 4 and 5. Therefore, the spectrograph broadening is not included in this version of PyISH. Future work to incorporate a UV IFS LSF is mentioned in Section {\ref{sect: discussion/future work}}.

Using the spectrum provided in Table 
\ref{table: user inputs, science}, we separate the continuum and emission lines by fitting a functional form of the continuum after masking out each of the emission lines that the user specifies. The mask at each emission line has a width of $\pm \lambda_{width}$ and is centered on each user-specified emission line wavelength. We  linearly interpolate using scipy's \texttt{interp1d} function over the masked-out emission lines to derive the flux contribution of the continuum only. The functional form is subtracted off of the original, user-provided spectrum to extract the flux contribution from each emission line as well. Overall, this process accounts for flux contributions from the user-specified spectrum, which includes emission lines and the continuum.

Using the extracted flux values, the module steps across the entire bandpass in unit steps equal to Equation \ref{equ: delta lambda}. At each step, the simulation has three paths for how the spatial maps are flux calibrated based on the user inputs described in Table \ref{table: user inputs, science}. First, if the wavelength step is not in the range of the user-provided spectrum, a map of zeros is created at that wavelength. Next, if the wavelength step sits within the wavelength range of the spectrum, but is not within the $\lambda_{width}$ of an emission line, then that wavelength step is attributed to continuum only. So, the spatial map for the continuum, labeled ``Continuum Map" in Table \ref{table: user inputs, science}, is calibrated to the continuum function at that step. This results in a three-dimensional cube of the continuum, with maps containing zeros where the spectrum does not exist, and flux calibrated continuum maps where the spectrum is specified. Lastly, if the step is at a wavelength within an emission line $\lambda_{width}$, the simulation will calibrate two different maps: a user-provided emission-line map and a user-provided continuum map. The continuum map is flux calibrated with the interpolated function of the continuum. The emission line map is scaled to the flux contribution for that respective emission line. 

It is important to note that at its current state, PyISH is equipped to handle spatial maps that can be calibrated to a single spectrum. Spatial maps corresponding to sources with unique spectra for each source are not currently in the scope of this simulation. Future discussion of this can be found in Section \ref{sect: discussion/future work}.

After stepping through the user-specified bandpass, the the result of the ``Image Processing" module described in Section \ref{subsect: Image Processing} is two three-dimensional input cubes. One is the cube of the continuum, with the user-specified continuum map flux calibrated. The other cube contains only the spatial maps from each respective emission line given in Table \ref{table: user inputs, science}, flux-calibrated according to Equation \ref{equ: delta lambda} across the specified $\Delta \lambda$. These 3D cubes, now in units of Rayleighs, serve as the input data for the UV IFS, that we will simulate through the optics specified in the following sections.

\subsection{Instrument Design Integration}
\label{subsect: telescope design integration}

Simulating a telescope and instrument design requires converting the on-sky image from a flux count, such as Rayleighs [$\frac{1}{4\pi} \cdot 10^{10} \text{photons} \cdot \text{s}^{-1} \cdot \text{m}^{-2} \cdot \text{sr}^{-1}$
] for our simulation, to a photon count. This depends on the number of reflections in the design, the coatings used on those reflective surfaces, the quantum efficiency of the detector, and the grating efficiency for each constituent of the grating array.  Thus, the photons we will receive from a given target depend on the assumed UV IFS instrument design. The efficiencies we are assuming for each optic are described in Table \ref{table: efficiencies} below.

\begin{table}[h!]
\begin{center}
 
\begin{tabular}{||c c c||} 
 \hline
 Label & Description & Source \\ [0.5ex] 
 \hline\hline
 QE & MCP Quantum Efficiency & SISTINE UV Rocket Instrument{\cite{Nell2024Far-Imaging}} \\ 
 \hline
  GE & Grating Efficiency & CHESS High Resolution Stellar Spectrograph \cite{Hoadley2016TheResults} \\
 \hline
  CE & Coating Reflectivity & GSFC XeLiF Coatings \cite{Quijada2025Far-ultravioletXeMgF2} \\ 
 \hline
\end{tabular}
\end{center}
\caption{Efficiencies assumed for a proposed HWO instrument layout.}
\label{table: efficiencies}
\end{table}

The SISTINE rocket payload was an imaging spectrograph designed as a prototype for the LUVOIR UV spectrograph, LUMOS {\cite{France2016TheMissions}}. SISTINE employed a cross-delay line microchannel plate (MCP) detector, a forerunner to the cross-strip design envisioned for HWO {\cite{France2017TheDesign}}. We assume the groove efficiency of the CHESS sounding rocket cross-dispersing grating {\cite{Tuttle2024UltravioletObservatory}}. The gratings and mirrors are covered with aluminum that has a Xenon-passivated Lithium Fluoride protective layer and are fabricated at GSFC and will be flown next year on the second flight of the INFUSE rocket payload {\cite{Haughton2025INFUSESpectrograph}}. They have shown peak reflectivities $>$ 0.8 in a laboratory environment {\cite{Quijada2025Far-ultravioletXeMgF2}}.

The telescope design is integrated using the HWO telescope PSF, as described in Section  \ref{subsect: Image Processing}. Currently, we assume a perfect PSF at the telescope focal plane, but that can be modified in the future to include surface roughness, telescope misalignment, and other sources of wavefront error. The user has the option to generate a new PSF at every wavelength step, or use one wavelength-specified PSF across the bandpass. The total efficiency of the telescope can be described by Equation \ref{equ: efficiencies}:

\begin{equation}
    \varepsilon_\lambda =  GE_\lambda * QE_\lambda * CE_\lambda^{N_{ref}}
    \label{equ: efficiencies}
\end{equation}

\begin{figure}[h!]
    \centering
    \includegraphics[width=0.95\linewidth]{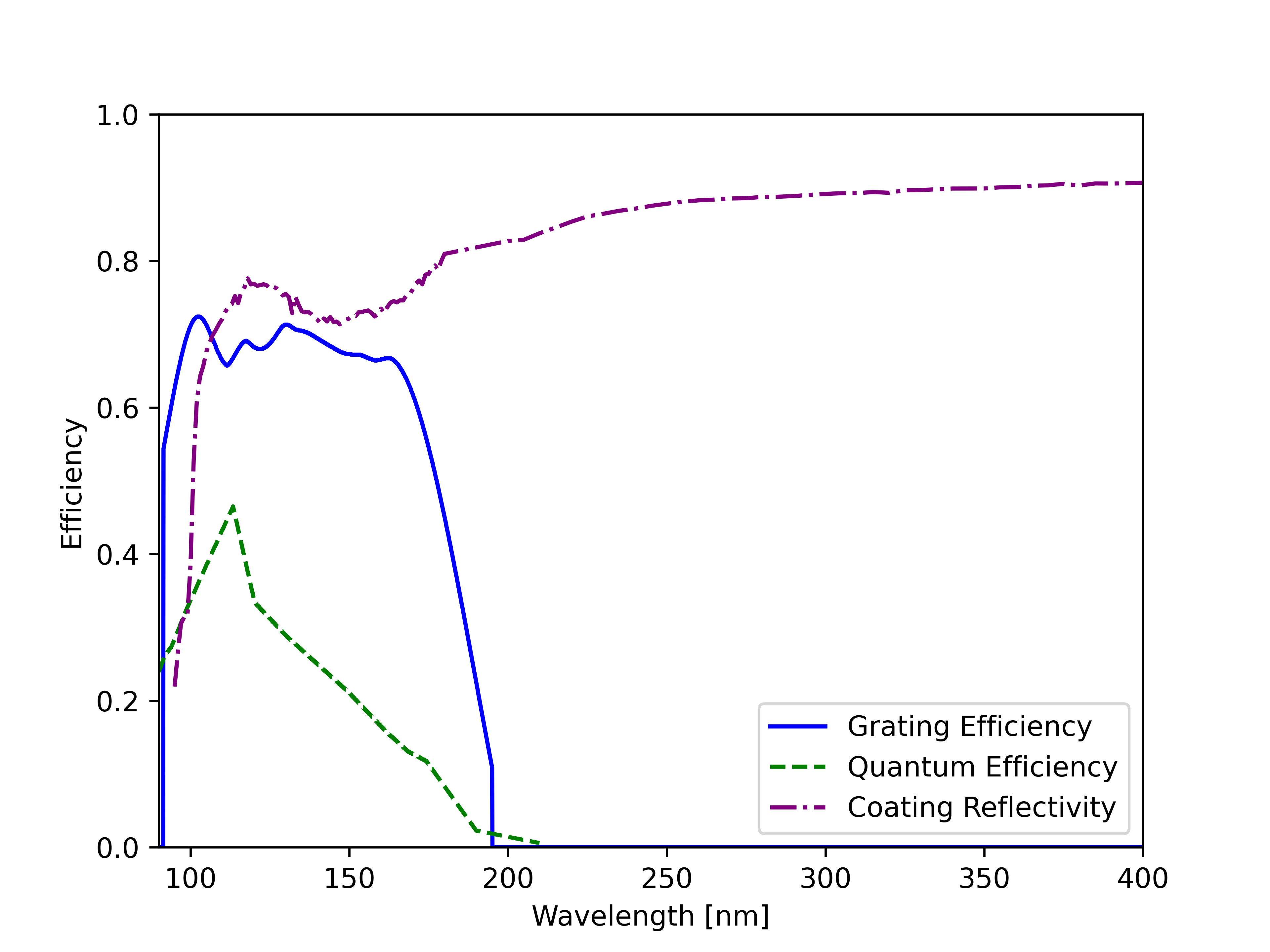}
    \caption{Efficiencies for each optical component assumed in the UV IFS simulation. The quantum efficiency is measured in flight, while the grating efficiency and coating reflectivity is measured in the lab {\cite{Nell2024Far-Imaging, Hoadley2016TheResults, Quijada2025Far-ultravioletXeMgF2}}.}

    \label{fig:efficiencies}
\end{figure}

$N_{ref}$ describes the number of reflections for the given design, as input by the user in Table \ref{table: user inputs, engineering}. $GE_\lambda$ is the grating efficiency, $QE_\lambda$ is the quantum efficiency of the detector, and $CE_\lambda$ is the coating reflectivity, each for a respective wavelength. Figure \ref{fig:efficiencies} shows the efficiency curves for each of the items listed in Table \ref{table: efficiencies}.
\newpage

The total efficiency from Equation \ref{equ: efficiencies} is included to convert the current flux calibrated three-dimensional cubes to photons. To do this, each step in the cube, representing a particular wavelength, is multiplied by the factor $\alpha$, in which

\begin{equation}
    \alpha_{\lambda} = \pi * (\frac{D_{ins}}{2})^{2}* \varepsilon_\lambda *T_{exp} * \Omega_{pix}^{2},
    \label{equ: telescope scaling}
\end{equation}

where $A_{geo}$ , $T_{exp}$, and $\Omega_{pix}$ are defined in Table \ref{table: user inputs, engineering}.

The FOV of the spatial map is not necessarily the same size as the user defined HWO IFS FOV, so the dimensions of each spatial map in the 3D cubes must be resized and zero-padded to fit in the particular section. We do the resizing in multiple steps. First, the slice is resized on the y-axis to $\phi$, given by Equation \ref{equ: as detector sees}, 

\begin{equation}
   \phi = \frac{FOV_{spectra} }{\Omega_{pix}}, 
   \label{equ: as detector sees}
\end{equation}

where $FOV_{spectra}$ and $\Omega_{pix}$ are given as inputs in Tables \ref{table: user inputs, science} and \ref{table: user inputs, engineering}, respectively. Resizing the y-axis to $\phi$ sizes the input science images to what would be seen on the HWO IFS detector. Next, if the $FOV_{IFS}$ is much larger than the input image, we zero-pad the y-axis, adding columns and rows of zeros iteratively to each side of the input image as if no photons were detected outside of the original image. For example, if the input science image is 2" x 2", but the UV IFS field of view (FOV) is 6" x 6", the original science image has to be resized so that the 2” x 2” FOV is correctly placed in the 6” x 6” FOV and padded with zeros. The reverse is also true; if the field of view for the spectra is much larger than the HWO IFS FOV, we iteratively cut pixels from the top and bottom of the spatial maps to size down to the HWO IFS FOV. 

A similar process is performed to resize the width of each spatial map. Equation \ref{equ: as detector sees} is implemented to resize the pixels on the x-axis to the length of $\phi$. Then, the length of the spatial maps are either padded with zeros, or pixels are deleted until the width matches the pixels in the HWO IFS detector. 

The result of ``Instrument Design Integration" module described in Section \ref{subsect: telescope design integration}
is the input three-dimensional cubes from Section \ref{subsect: Image Processing}, with each spatial map in the cube converted to a photon count and properly sized to the user-specified IFS FOV depending on the optical design of the telescope and instrument.

\subsection{IFS Slicer}
\label{Subsect: IFS slicer}

Next, we implement the IFS slicer optic. The assumed slicer is reflective, consisting of many small mirrors on a single piece of material, tilted in slightly different directions. To simulate the effect of this optic, the spectral dimension of the two three-dimensional data cubes from Section \ref{subsect: Image Processing} are once iterated over in unit steps equal to Equation \ref{equ: delta lambda}. In each step, the spatial map is sliced into many smaller images pieces consistent with the number of slices in the IFS slicer, defined as $N_{slices}$ in Table \ref{table: user inputs, engineering}. 
The result is two, three-dimensional input cubes as before, but now spatially separated into many narrow slices. This is equivalent to having many narrow three-dimensional cubes, each corresponding to one slice of the IFS. In the following section, the spectra from the input data cubes will be placed onto the detector, then reconstructed to be the final data as seen by a UV IFS.

\subsection{Detector Readout and Final Products}
\label{subsect: detector readout/final products}

\begin{figure}[h!]
    \centering
    \includegraphics[width=0.5\linewidth]{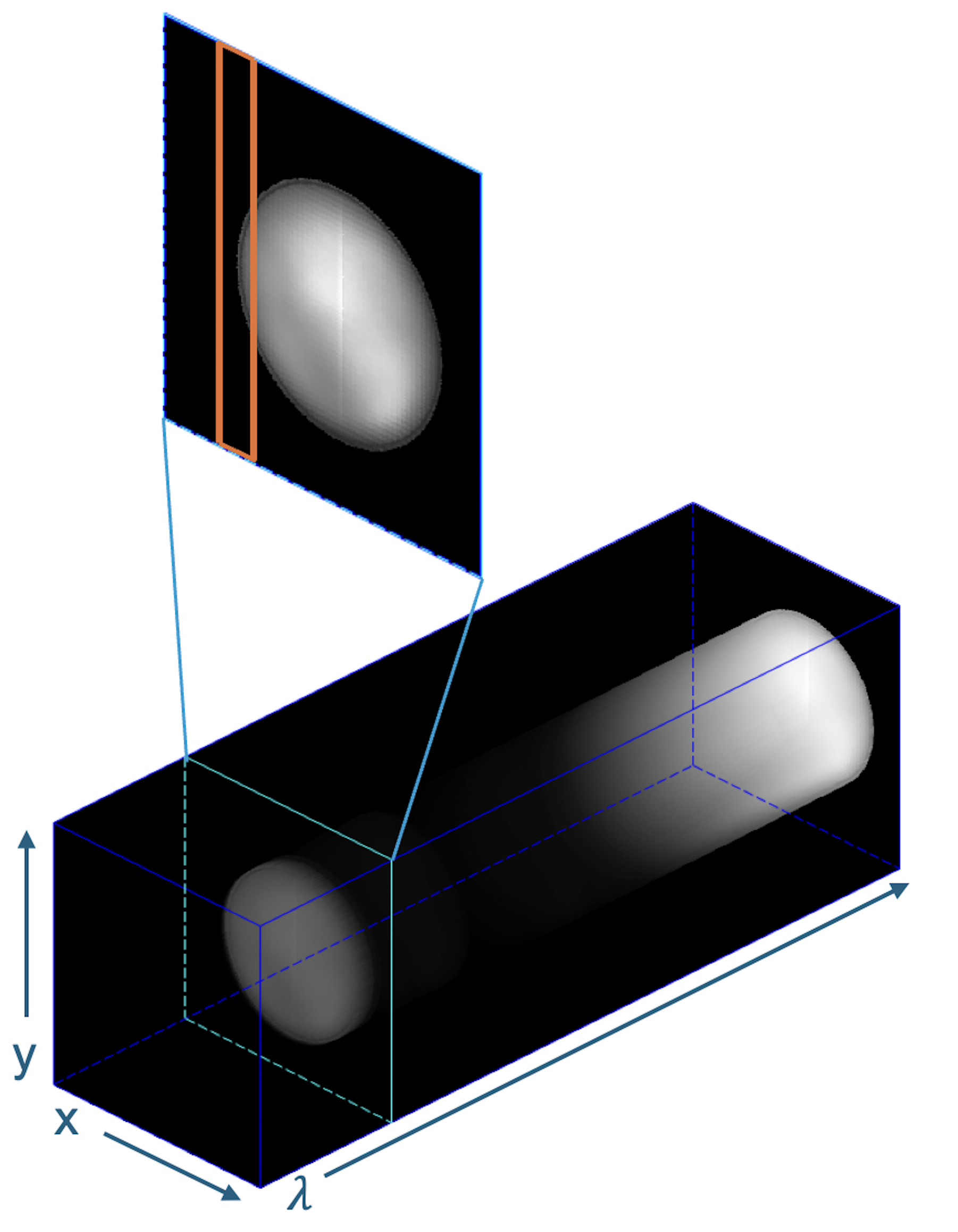}
    \caption{Example of choosing the correct data to insert at a particular pixel in the detector. The IFS slice [shown in orange] corresponds to a particular section of the detector, as the number of detector sections is dependent on the number of IFS slices. The spatial map [boxed in blue] corresponds to a particular wavelength, which can be traced to a pixel within a detector section. }
    \label{fig:slice to insert}
\end{figure}

The goal of this module is to place the spectra, now in units of photon counts, onto the detector, and use post-processing reconstruction to output a final three-dimensional data cube. As described in Section \ref{subsect: integration user inputs}, the detector on an IFS must be sectioned into areas for the spectra for each IFS slice, in order to maintain spatial information. The simulation steps through each section of the detector. For every section, the x-axis is as long as the desired bandpass from Section \ref{subsect: integration user inputs}, in units of pixels. Each pixel across the x-axis corresponds to a specific wavelength in the bandpass. As we assume a Nyquist-sampled spectrum, each wavelength step consists of two pixels. To place the correct data in the section, we identify the spatial map corresponding to the pixel, and the slice corresponding to that section of the detector. A depiction of this process is shown in Figure \ref{fig:slice to insert}.

 \newpage

While the projected slice width on the detector may span multiple detector resolution elements, we assign each original slice to be two pixels wide in the reconstructed final data cube. The simulation sums the selected data along the x-axis, so that it is one pixel wide. This produces a one-dimensional spectrum, equivalent to physically dispersing the slice as a long-slit spectrum. Then, we concatenate two copies of the data together, such that it is as tall as the section, and two pixels wide. This process of choosing the associated slice within the 3D continuum and placing it in the correct space is repeated for every pixel in a given detector section. Then, every section is iterated until the continuum is inserted in every detector section associated with an IFS slice. A flowchart describing this process is shown below in Figure \ref{fig: detector section flowchart}. 
\newpage
\begin{figure}[h!]
    \centering
    \includegraphics[width=0.9\linewidth]{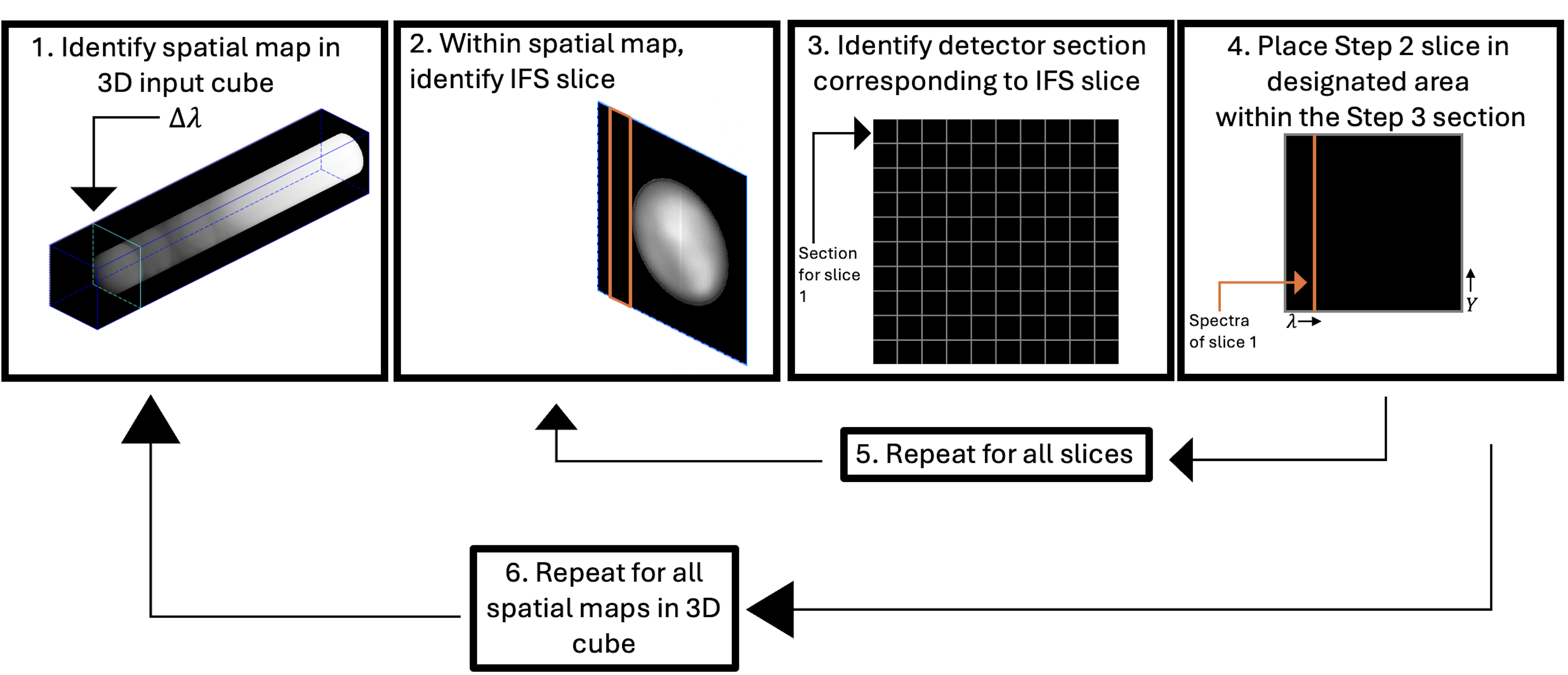}
    \caption{Flowchart describing the process of selecting an IFS slice from a particular spatial map, and placing the data in the appropriate section of the detector. This process is used to record the continuum of a source on the detector.}
    \label{fig: detector section flowchart}
\end{figure}

The flux from each emission line is added to an empty copy of the HWO detector. Rather than moving across every pixel in a given section, the spatial maps containing values between $\lambda_{width}$ for each emission line are added using Equation \ref{equ: emission line placement}, 

\begin{equation}
    \beta = \frac{\lambda - \lambda_{start}}{\Delta\lambda} * 2 ,
    \label{equ: emission line placement}
\end{equation}

where $\beta$ is the pixel that the emission line of wavelength $\lambda$ will be placed, $\lambda_{start}$ is the beginning of the bandpass defined in Table \ref{table: user inputs, engineering}, and the factor of two accounts for Nyquist sampling. Whereas the continuum is placed by iterating through each pixel, and identifying the corresponding slice, the emission line spectra are placed using the selected pixels as determined by Equation \ref{equ: emission line placement}.  Following the same process described for the continuum, the emission line image is resized and placed at $\beta$ and $\beta + 1$, again accounting for Nyquist sampling.

The two arrays that represent the detector footprint are added together, and dark current and Poisson shot noise are simulated. Dark current is included by adding the constant rate specified by Table \ref{table: user inputs, engineering} to each detector section. Then, Poisson shot noise is added to each section using numpy's \texttt{random.poisson} function. Other sources of noise beyond Poisson shot noise and dark current are omitted from this version of PyISH. Inclusions of additional noise terms are discussed in Section {\ref{sect: discussion/future work}}.

The user has three potential data products from the simulation. The first is a FITS file of the full, not-post-processed, two-dimensional detector. This is also available as a three-dimensional FITS file of each section in the detector. The x and y axes of the 3D FITS file are the dimensions of each section in the detector, and the z axis iterates over the sections corresponding to how the slices were cut in the spatial maps from left to right. The second is a three dimensional cube for the user's specific science case with their input architecture. We do this by moving section by section, iterating through each pixel and joining the strips for each respective pixel into a spatial map for that step in the bandpass. This is output as a FITS file. The third data product is available as an option at the conclusion of the simulation, to select a spectrum from a particular spaxel. The user specifies which coordinate to derive the spectrum, and the simulation will step through the the bandpass in unit steps corresponding to Equation \ref{equ: delta lambda}, selecting the photon count associated with that coordinate at every step to construct a spectrum from that area.

\section{Simulation Tool Example: Europa}
\label{sect: europa example}

The science case, ``Do large icy bodies and moons harbor habitable environments in their interiors?" by Richard Cartwright et. al. was chosen to be simulated as an example{\cite{Cartwright2025AssessingHWO}}. This science case focuses on the potentially habitable ocean worlds in the outer solar system. Because Europa has a long record of existing FUV observations collected by HST, it was chosen as the target of interest for the simulated UV IFS. Europa is a moon of Jupiter that likely has a subsurface saline ocean that may be geologically connected to its surface {\cite{Daubar2024PlannedMission}}. Compelling evidence for such geologic communication was potentially detected in form of concentrated H and O emission lines detected above Europa's limb along its south polar region, representing a possible $H_2O$ vapor-dominated plume {\cite{Roth2014TransientPole}}. The subsections that follow will leverage this HST data to simulate a UV IFS view of Europa with the upcoming HWO.

\subsection{Integration of User Inputs}
\label{subsec: integration europa}

The science inputs specified in Table \ref{table: user inputs, science} are described in Section \ref{subsec: image processing europa}. The engineering inputs for the Europa science case are described below in Table \ref{table: user inputs example}.

\begin{table}[h!]
\begin{center}
\begin{tabular}{||c c||} 
 \hline
  Input & Description \\ [0.5ex] 
 \hline\hline
  $N_{Slices}$ & 100 \\ 
  \hline
 $dim_{detector,1}$ & 94 [mm] \\
 \hline
$dim_{detector,1}$ & 94 [mm] \\
 \hline
 $\omega_{dark}$ & 4.4 [cts/cm$^2$/s] \\
 \hline
$\lambda$ & 121.6 [nm] \\
 \hline
 $R_{\lambda}$ & 1500 \\
 \hline
 $bp$ & 100 [nm]\\
 \hline
 $FOV_{IFS}$ & 1 [as] \\
 \hline
 $\Omega_{pix}$& 6  [mas/pixel] \\
 \hline
 $\gamma$ & 10 $\mu m$/pixel \\
 \hline
 $D_{ins}$ & 6.5 [m] \\
 \hline
$T_{exp}$ & 50000 [sec] \\
 \hline
$\lambda_{start}$ & 100 [nm] \\
 \hline
 $N_{ref}$ & 4\\
 \hline
 $EAC$ & 1  \\ [1ex] 
 \hline
\end{tabular}

\end{center}
\caption{Example UV IFS engineering parameters use to simulate Europa's disk and limb plume.}
\label{table: user inputs example}

\end{table}

We chose a detector of size 94 mm x 94 mm based on the INFUSE design{\cite{Haughton2025INFUSESpectrograph}}. The pixel pitch and plate scale were chosen to be in the range desired for the HWO UV IFS according to submitted SCDDs{\cite{Scowen2025TechnologyObservatory}}. $\omega_{dark}$, the dark current rate, was chosen based on previous measurements during commissioning of UV solar system instruments{\cite{Davis2025JupiterManeuver}}. The minimum field of view and resolution were chosen as low end constraints that we believed would be achievable given the detector parameters. The number of slices, $N_{Slices}$ was chosen to be the same order of magnitude as current state-of-the-art reflective slicer capabilities{\cite{Sukegawa2023Ultra-compactIFU}}. The inscribed diameter was chosen based on EAC 1, and the exposure time was chosen as a conservative estimate{\cite{Feinberg2024TheOpportunities}}. The number of reflections assumes two reflective surfaces after the primary and secondary mirrors, similar to INFUSE (see Section \ref{subsec: telescope europa}). 

It is important to note that while the UV IFS is the new instrument addition in EACs 4 and 5 (mentioned in Section {\ref{sect:intro}}), neither telescope nor instrument specifications have been released to the public. Therefore, the primary mirror diameter, reflections, and efficiencies in the current example are tuned to EAC 1. The specifications are also available for EACs 2 and 3, which can serve as inputs to PyISH. Once available, PyISH will be able to run on all current and future EACs.

\begin{figure}[h!]
    \centering
    \includegraphics[width=0.6\linewidth]{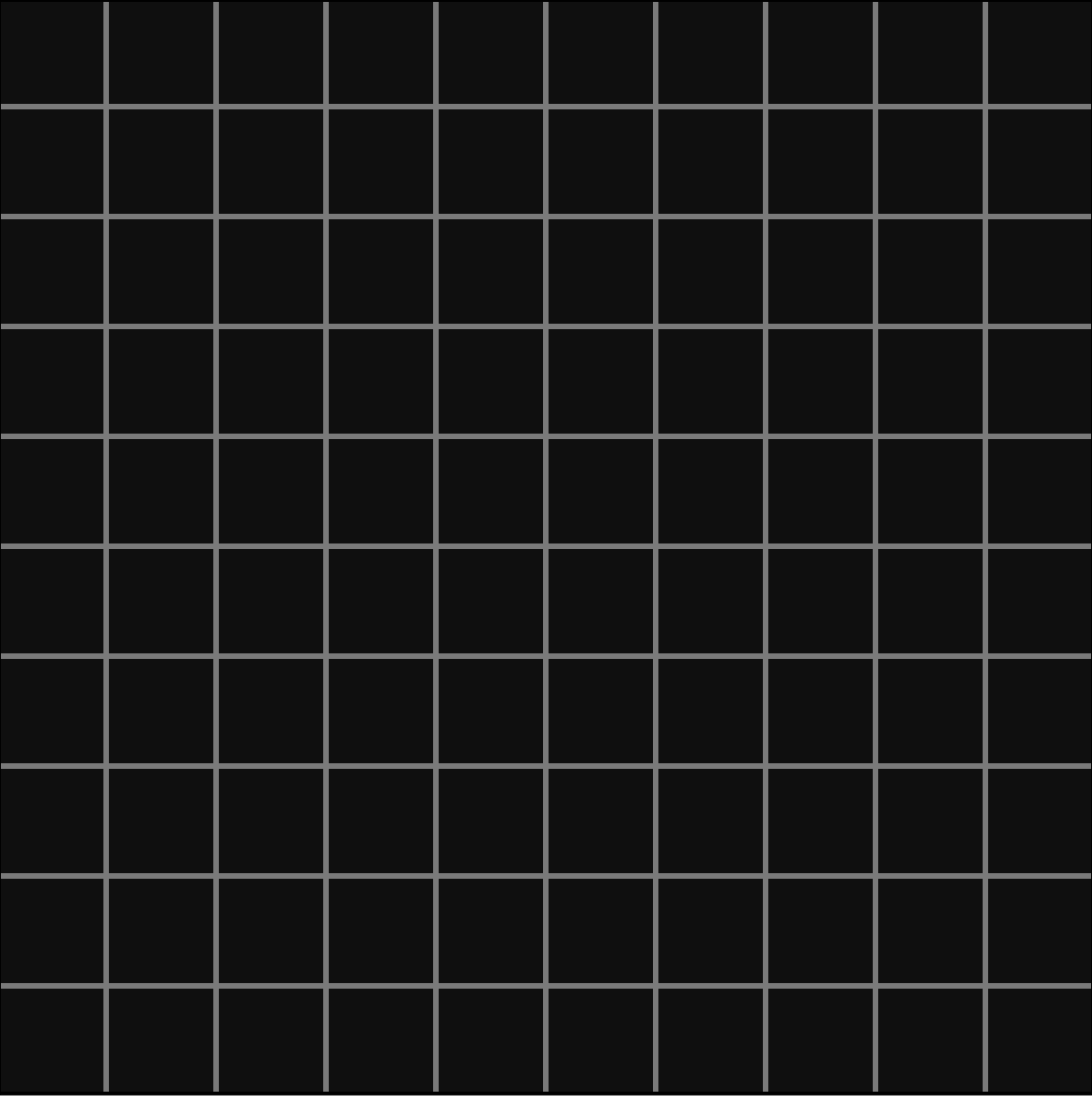}
    \caption{Empty detector split with the same number of sections in x and y. In this example, the detector has 10x10 sections. }
    \label{fig:10x10sections}
\end{figure}

The process described in Section \ref{subsect: integration user inputs} is implemented using the parameters in Table \ref{table: user inputs example}. The detector is first sectioned into even cuts on the x and y axes resulting in 100 total sections, equal to the number of specified IFS slices. The empty detector with these sections is shown in Figure \ref{fig:10x10sections}.

\newpage

After determining the minimum detector size needed using Equations \ref{equ: min x detector} and \ref{equ: min y detector}, we found that the current organization of detector sections will not fit on the detector size we specified. Specifically, the height of each section will meet the minimum height needed to achieve the FOV, as determined by Equation \ref{equ: min y in pixels}, but the width of each section as it is currently split in the detector will not meet the minimum width to achieve a bandpass of 100 nm with spectral resolving power of 1500, as determined by Equation \ref{equ: min x in pixels}. Therefore, Option B is invoked to reorganize the sections on the detector, in which we remove columns and add rows, making the sections wider and shorter. After the reorganization, the sections are the correct size, and can all fit on the given detector space. A figure of the new, reorganized sections is depicted below in Figure \ref{fig:detector sections update example}.

\begin{figure}[h!]
    \centering
    \includegraphics[width=0.6\linewidth]{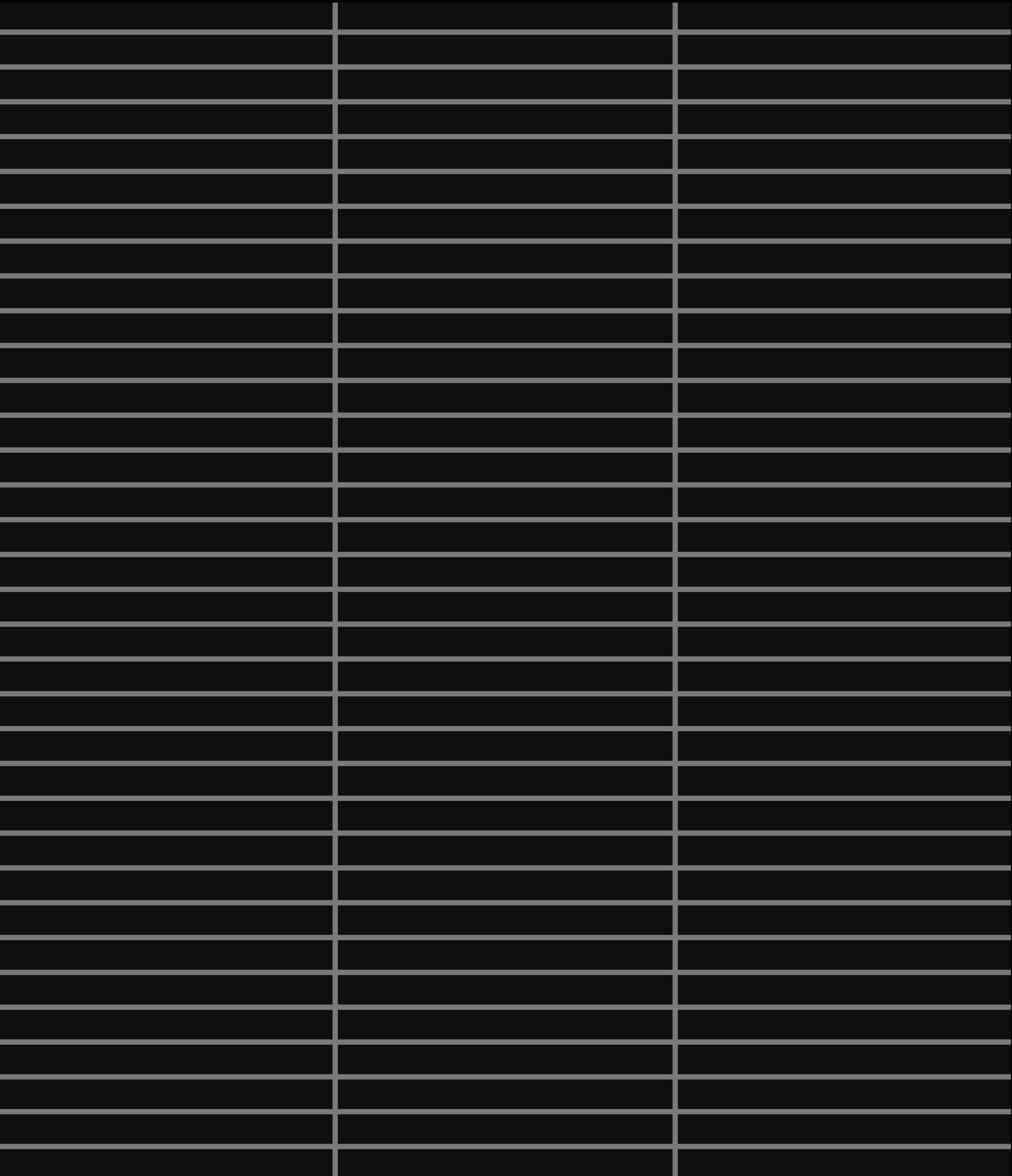}
    \caption{Empty detector split into 100 sections after reorganization according to Section \ref{subsubsect: option B: Adjust Width}, in which more space was required on the detector's x-axis to account for spectral constraints. This is the detector organization used in the current example.}
    \label{fig:detector sections update example}
\end{figure}

\subsection{Update Parameters}
\label{subsec: update parameters europa}

We choose to use the entire detector space of 94 mm x 94 mm specified for this example. This gives an updated FOV and bandpass shown in Table \ref{table: updated science}.

\begin{table}[h!]
\begin{center}
\begin{tabular}{||c c||} 
 \hline
  Name & Updated Science Parameter \\ [0.5ex] 
 \hline\hline
  $FOV_{IFS, updated}$ & 1.66 arcseconds x 1.66 arcseconds \\ 
  \hline
 $bp_{IFS}$ &   126.99 nm  \\
 \hline
 Minimum Detector Size & 56.78 mm x 74.04 mm\\
 \hline
\end{tabular}
\end{center}
\caption{Updated IFS science constraints after deciding to use the entire space on the detector.}
\label{table: updated science}

\end{table}

As shown, even though we asked for a field of view of 1 arcsecond x 1 arcsecond, we had extra space on the detector to optimize, leading to a field of view of 1.66 arcseconds by 1.66 arcseconds. In addition, the bandpass is increased by 26.99 nm. 

\subsection{Image Processing}
\label{subsec: image processing europa}

Table \ref{table: user inputs, science, europa} below specifies the science inputs utilized in this module.

\begin{table}[h!]
\begin{center}
\begin{tabular}{||c c||} 
 \hline
  Science Input & Description \\ [0.5ex] 
 \hline\hline
  Name & Ly-$\alpha$, O I, [O I] \\ 
  \hline
 $\lambda_{emission}$ &121.6 nm, 130.4 nm, 135.6 nm  \\
 \hline
 $\lambda_{width}$ & 4 nm, 4 nm, 5 nm \\
 \hline
 Spatial Maps & See Figure \ref{fig: europa maps} \\
 \hline
 Spectrum & See Figure \ref{fig:emissionvscont}\\
 \hline
 Continuum Map & See Figure \ref{fig: europa maps} \\
 \hline
 $FOV_{spectra}$ & 1.60 arcsec \\ [1ex] 
 \hline
\end{tabular}
\end{center}
\caption{Example UV IFS simulation engineering constraints set by the user. Broad slits in the STIS observations contribute to the large $\lambda_{width}$ values compared to the intrinsic $\lambda_{width}$ of each emission line. }
\label{table: user inputs, science, europa}
\end{table}

\begin{figure}[h!]
    \centering
    \includegraphics[width=0.9\linewidth]{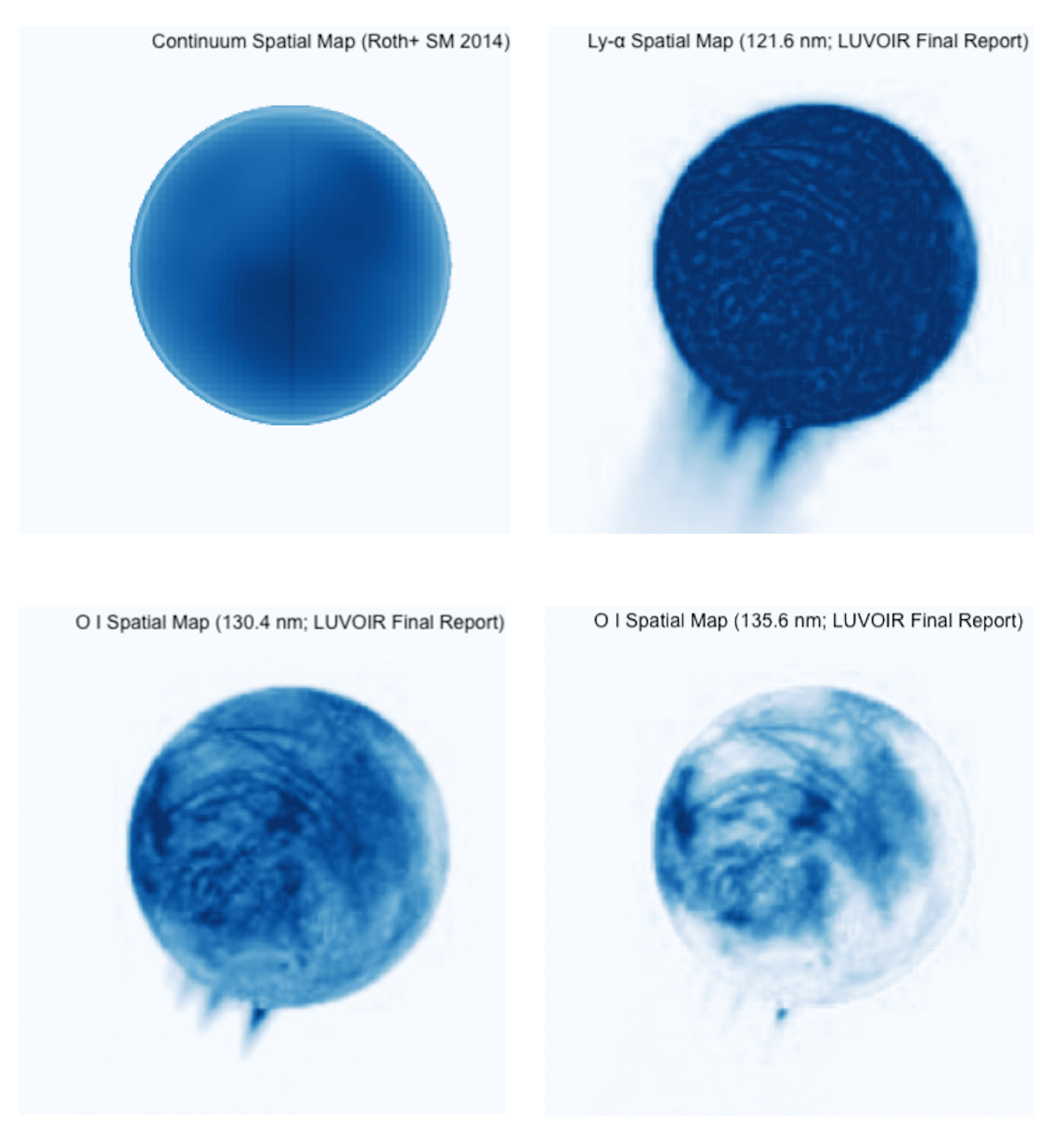}
    \caption{Spatial maps input for the Europa science case example, depicting three separate emission lines and a modeled surface reflectance image.}
    \label{fig: europa maps}
\end{figure}

Three emission lines in the UV were identified in an HST STIS study of Europa (Ly-$\alpha$ and two OI lines){\cite{Roth2014TransientPole}}. High angular resolution spatial maps of those lines were simulated for the LUVOIR architecture study, which we use here{\cite{ScienceandTechnologyDefinitionTeam2019LUVOIRREPORT}}. We use the model of the UV continuum from the HST STIS study of Europa{\cite{Roth2014TransientPole}}. The emission line spatial maps and surface continuum map are depicted in Figure \ref{fig: europa maps}. 

\newpage

After convolving with the PSF, each map was flux calibrated using the user-provided spectrum from Roth+ 2014 {\cite{Roth2014TransientPole}}. A representation of the continuum was created following Section \ref{subsect: Image Processing} and the information provided in Table \ref{table: user inputs, science, europa}. The functional form of only the continuum, and the flux of the entire spectrum is shown below in Figure \ref{fig:emissionvscont}.
\newpage
\begin{figure}[h!]
    \centering
    \includegraphics[width=0.8\linewidth]{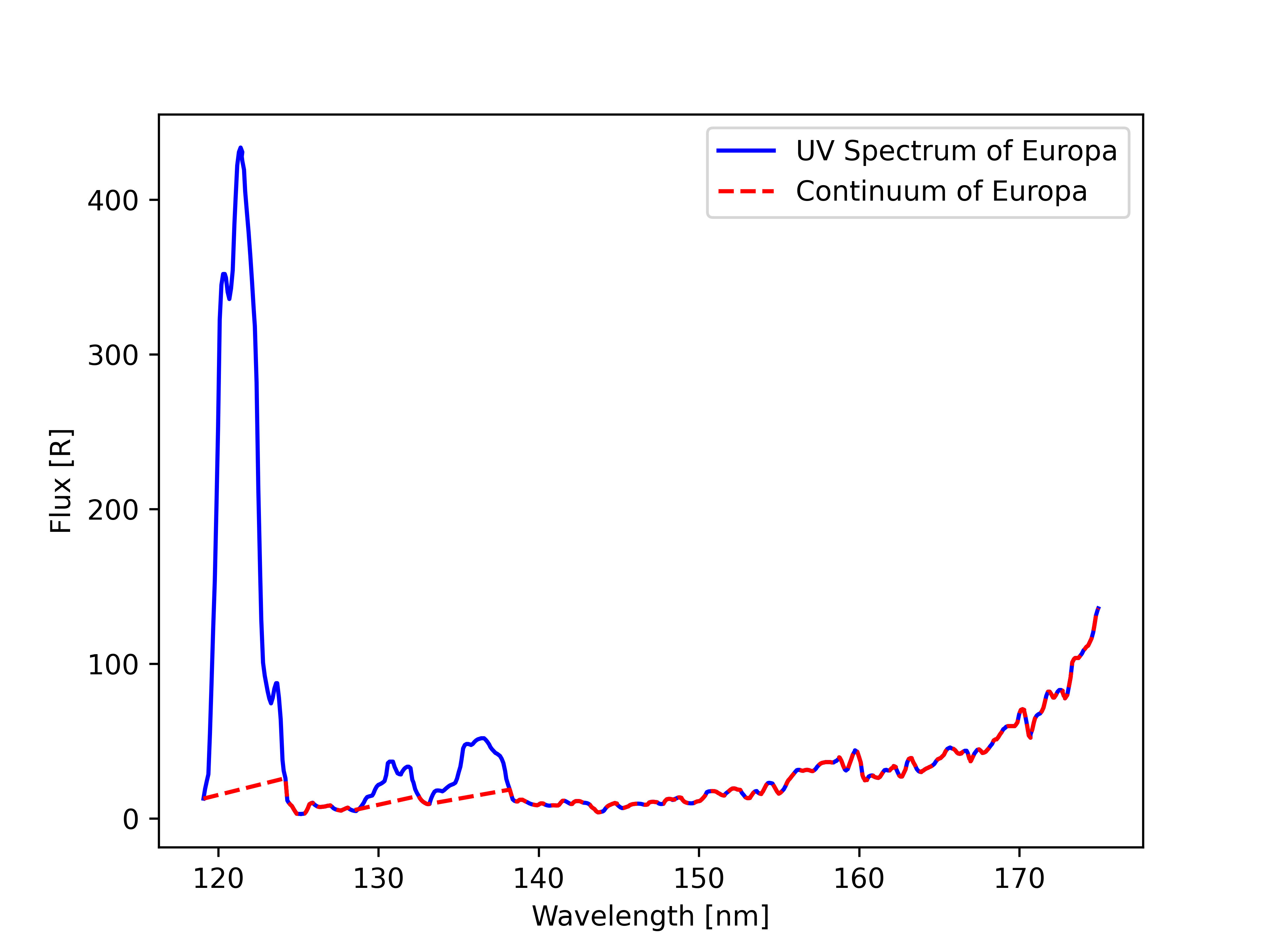}
    \caption{UV spectrum of Europa, along with the functional form of the continuum extracted by masking the emission lines (Input spectrum from Roth et al., 2014{\cite{Roth2014TransientPole}}).}
    \label{fig:emissionvscont}
\end{figure}

The flux of the spectrum is in blue, and the flux due to the continuum is dashed in red. With the emission and continuum extracted, the spatial maps are flux calibrated in units of Rayleighs. The spatial maps of the emission lines are calibrated to the area between the dashed red line and solid blue line, while the spatial map of the continuum is flux calibrated across the entire bandpass to the area under the dashed red line.

\subsection{Instrument Design Integration}
\label{subsec: telescope europa}

As mentioned in Section \ref{subsect: telescope design integration}, the goal of this module is to take the input three-dimensional cubes, and assume an instrument design to correctly calibrate the cubes to a photon count, as well as appropriately size the input cubes to the instrument's field of view. The instrument design we chose has 4 reflections (two in the telescope and two in the IFS), as input in Table \ref{table: user inputs example}. Every map is multiplied by the factor from Equation \ref{equ: telescope scaling} at each wavelength to convert the flux calibrated cubes to photons.

In addition, the 3D cubes are resized according to the FOV determined in Equation \ref{equ: hwo FOV}. This is done by first converting the original pixels to the size of the pixels on an HWO IFS detector using Equation \ref{equ: detector in pixels}, then either padding with zeros or removing pixels according to the HWO IFS FOV determined in Equation \ref{equ: hwo FOV}. Overall, this results in 3D cubes that are exactly the size of the HWO IFS FOV, with the same plate scale, and in units of photon counts.

\subsection{IFS Slicer}
\label{subsec: slicer europa}

Every map, now in units of photon counts following Section \ref{subsec: telescope europa}, is sliced according to the user input in Table \ref{table: user inputs, engineering} for the number of IFS slices. In this example, this is 100 slices. The spatial maps are sliced using  Astropy's \texttt{Cutout2D} function{\cite{TheAstropyCollaboration2022ThePackage}}. An example of one of these cutouts is shown below in Figure \ref{fig:cutout example}. 
\newpage
\begin{figure}[h!]
    \centering
    \includegraphics[width=0.9\linewidth]{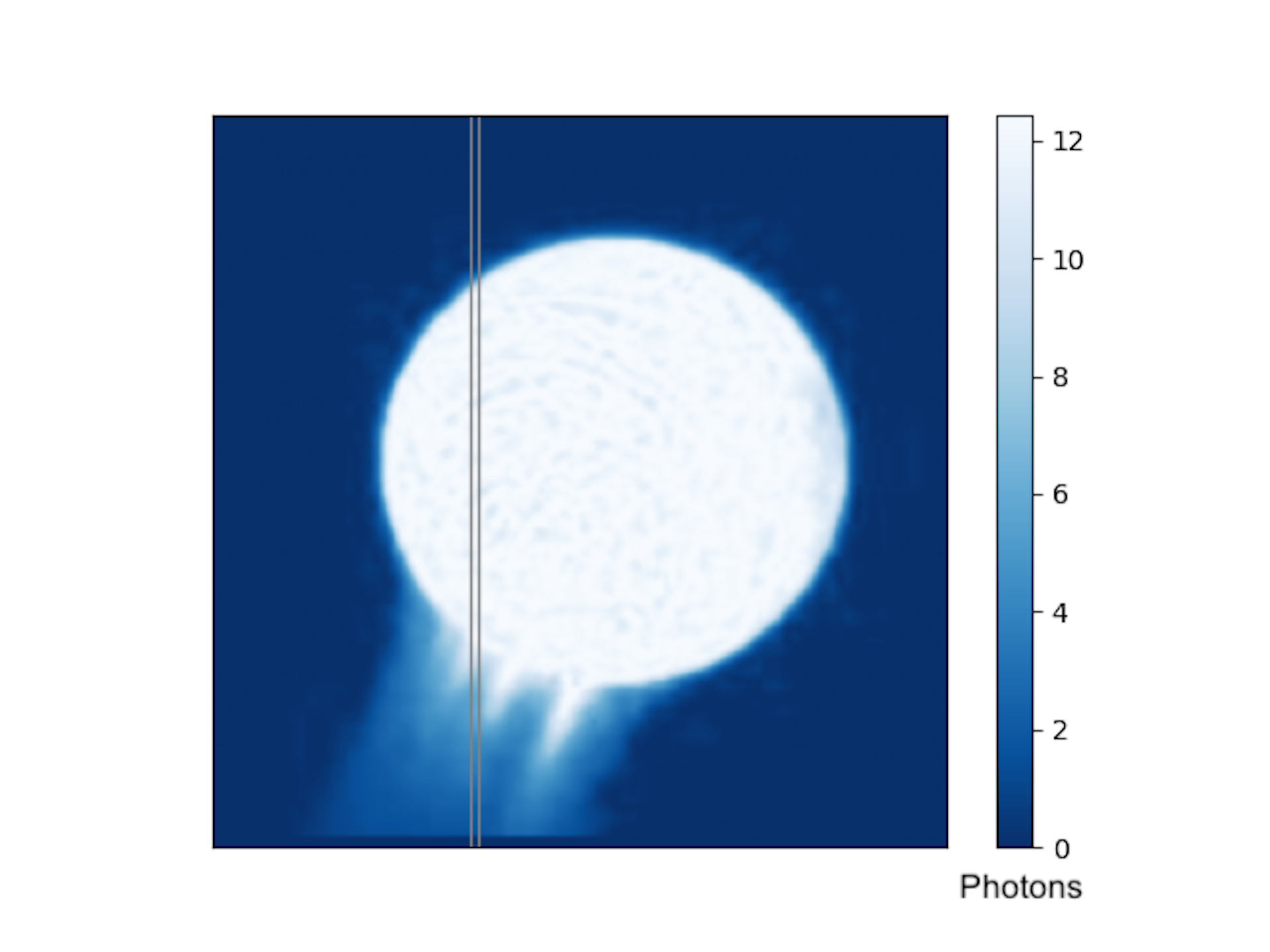}
    \caption{One slice of Europa after the IFS simulation's ``IFS Slicer" module, with the spatial map in units of photons and the FOV of the specified IFS simulation. As specified by Table \ref{table: updated science}, the image is 1.66 arcseconds across. Europa's disk is 1.02 arcseconds across.}
    \label{fig:cutout example}
\end{figure}

The cutouts are taken left to right, such that the first IFS cutout will be the leftmost slice, while the last cutout is the rightmost. The result of this module is cutouts for every spatial map along the three dimensional cubes for both the emission lines and the continuum. 

\subsection{Detector Readout and Final Products}
\label{subsec: readout europa}

Finally, the slices from 3D cubes pertaining to the continuum and emission lines of Europa are placed in the sections of the detector according to Section \ref{subsect: detector readout/final products}. As mentioned previously, the continuum is added to the HWO detector section by section, by iterating through each pixel on the x-axis. The emission lines are added onto an empty copy of the HWO detector, placed according to Equation \ref{equ: emission line placement}. The two detectors are added together, the dark rate specified by Table \ref{table: user inputs, science, europa} is added, and Poission shot noise is included. 

Once the data has been inserted into the detector, post processing is done to retrieve data products for the user. The first potential data product is a stitched .FITS file of every section of the detector. This is shown below in Figure \ref{fig:stitcheddetector}. 

\begin{figure}[h!]
    \centering
    \includegraphics[width=0.95\linewidth]{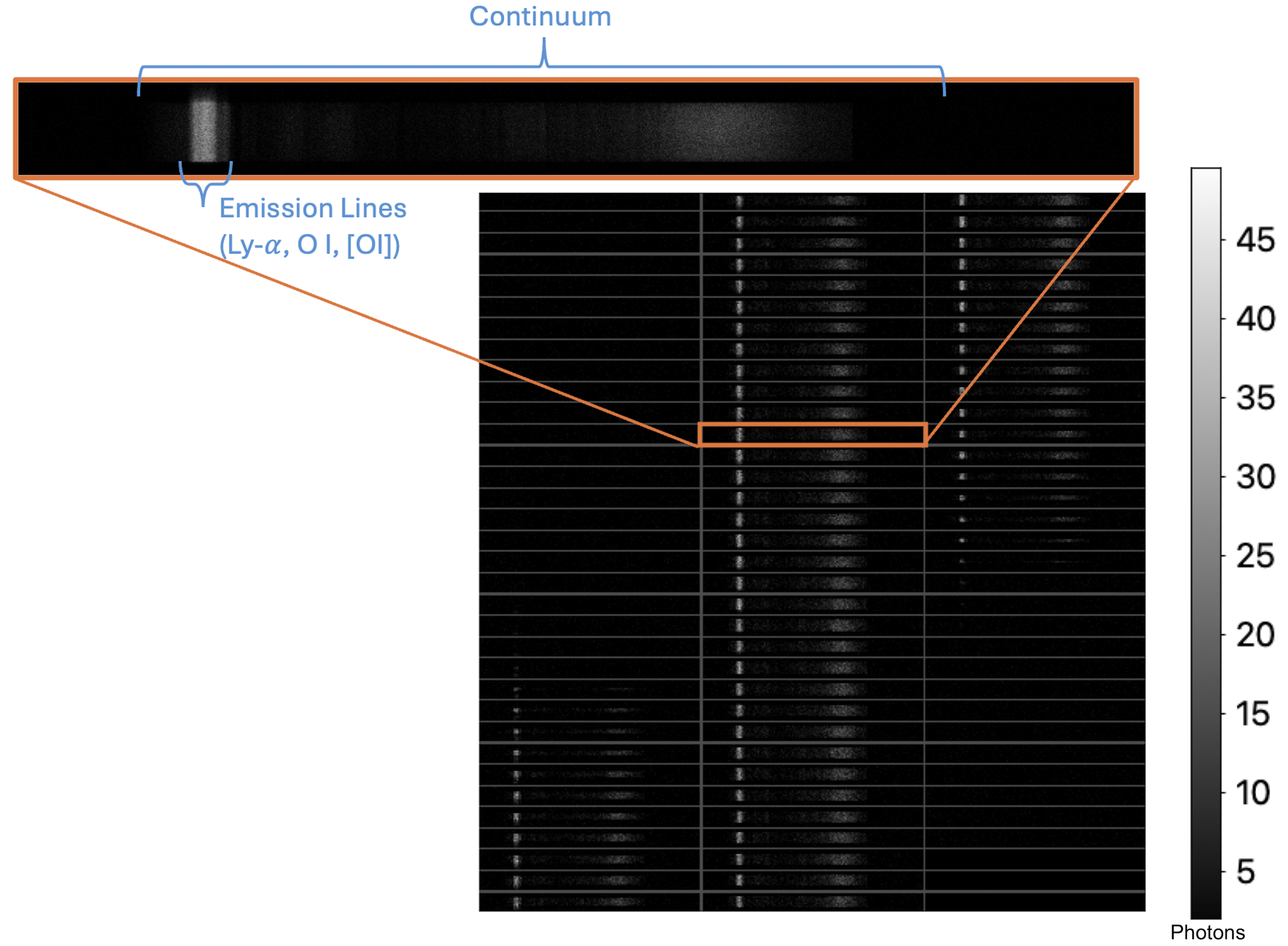}
    \caption{Detector from Figure \ref{fig:detector sections update example}, now filled with spectra specified in Section \ref{subsec: image processing europa} of Europa.}
    \label{fig:stitcheddetector}
\end{figure}

The IFS slices iterate vertically from top to bottom, then horizontally from left to right; the first IFS slice of the spatial maps is located in the top left, and the last IFS slice has its spectra in the bottom right. As mentioned in Section \ref{subsect: detector readout/final products}, the detector sections are also available in a three-dimensional cube format, in which iterating through the z-axis iterates through the spectra of each slice.

The second potential data product is a three dimensional cube. The cube is reconstructed as described in Section \ref{subsect: detector readout/final products}, and provided as a .FITS file. An example of the three dimensional cube for two exposure times of Europa is provided in Figure \ref{fig: europa maps and cube} below.

\begin{figure}[h!]
    \centering
    \includegraphics[width=0.9\linewidth]{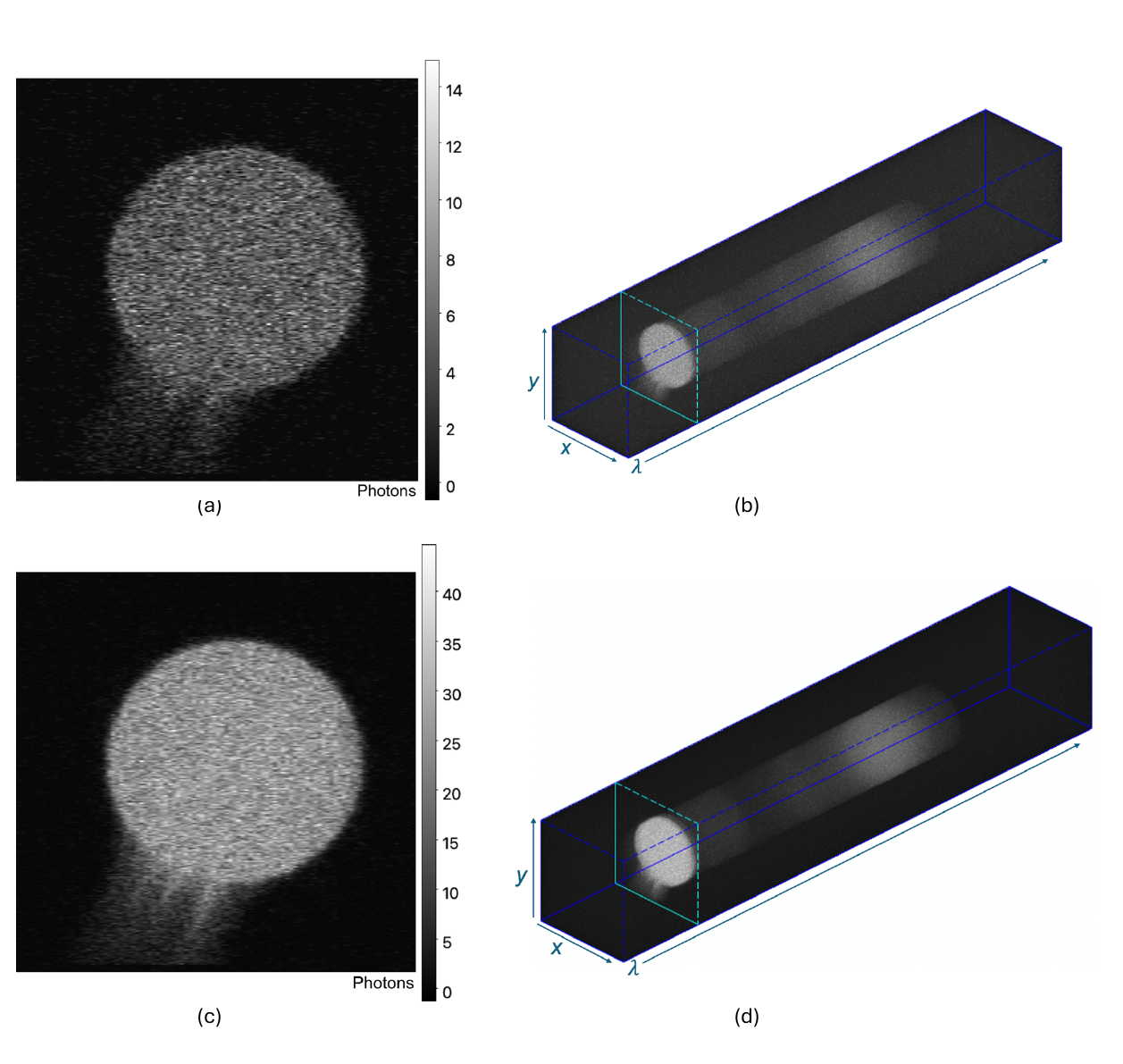}
    \caption{Emission line spatial maps and 3D cubes of different exposure times for the same science case, as it would be depicted by an HWO UV IFS with constraints according to Table \ref{table: updated science}.}
    \label{fig: europa maps and cube}
\end{figure}

(a) and (b) in Figure \ref{fig: europa maps and cube} depict the Lyman-$\alpha$ spatial map and the entire three-dimensional cube, respectively, for an exposure time of 10,000 seconds. (c) and (d) in Figure \ref{fig: europa maps and cube} show the same information for an exposure time of 50,000 seconds. As expected, a higher exposure time results in more photons, yielding a spatial map with a higher signal-to-noise ratio.

The third potential data product is provided as an option at the end of the simulation, in which the user can select a certain spaxel, and receive the spectrum from that spaxel. A depiction of this is below in Figure \ref{fig:spaxalspectrum}.

\begin{figure}[h!]
    \centering
    \includegraphics[width=0.95\linewidth]{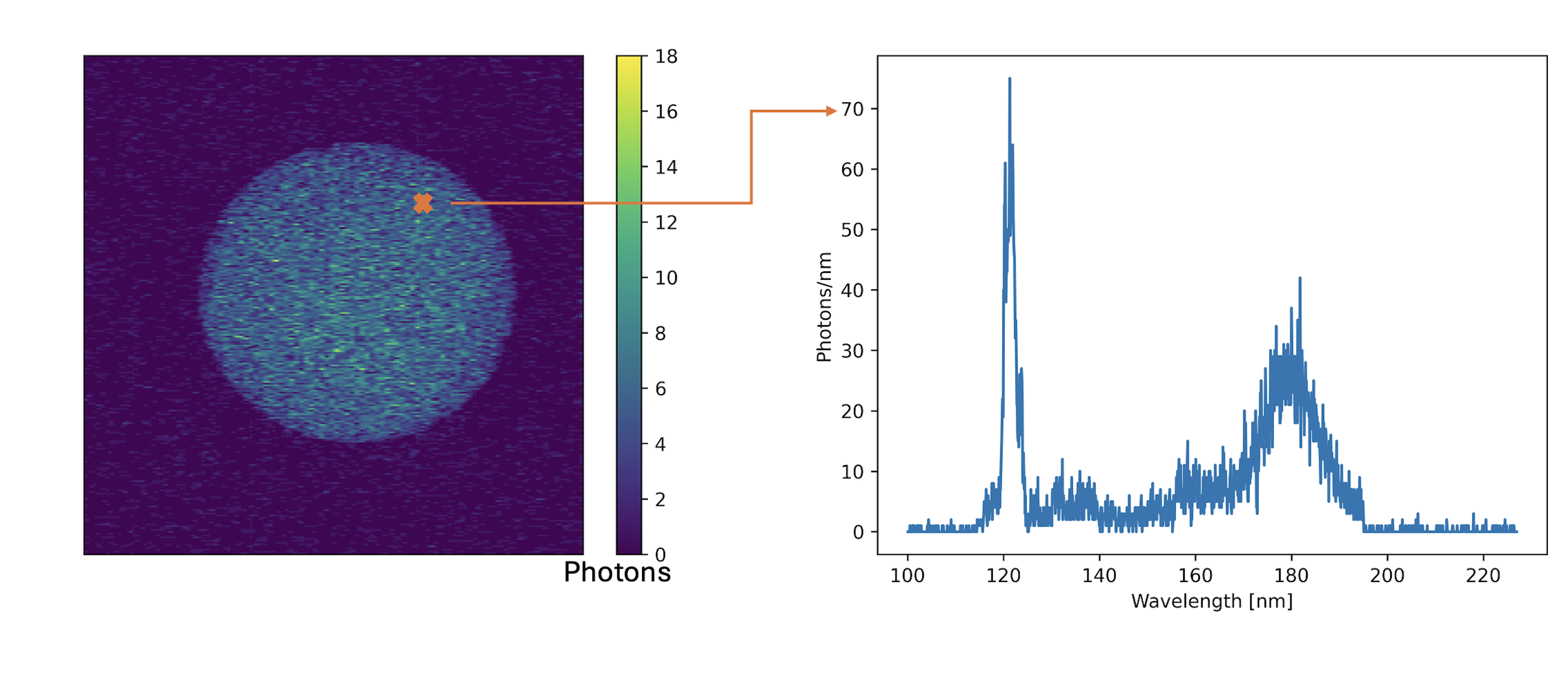}
    \caption{Extracted spectrum from a particular spaxel in the HWO UV IFS simulated 3D cube.}
    \label{fig:spaxalspectrum}
\end{figure}

For the spaxal highlighted in orange, the spectrum across the entire bandpass is extracted on the rightmost side of the figure.

\section{Discussion and Future Work}
\label{sect: discussion/future work}

The simulation tool we developed allows the community to showcase the power of a HWO-like UV IFS by simulating science cases and trading potential instrument parameters. With a spectrum, high resolution maps of the science case, and a completed table of the desired instrument parameters, scientists can simulate any science case they desire. With potential instrument designs, engineers can trade the most effective setup to encompass the best performing instrument for the community, taking in both science requirements such as spectral resolving power and field of view, and engineering requirements such as the type/size of the detector(s), gratings, reflective slicers, and coatings.

This tool can be utilized to understand how a UV IFS can capture a specific science case. The example provided uses inferred instrument designs, and optical properties are assumed using the engineering specifications for potential telescope, instrument, and detector properties for the Habitable Worlds Observatory, provided by the HWO TMPO. Additional reasoning for the engineering choices we made in the example are in Sections \ref{subsec: integration europa} and \ref{subsec: telescope europa}.

For the advanced user, this tool can be used to inform optical designs to understand the science gains achievable by trading different detector parameters, reflective slicer parameters, as well as coatings and gratings. Although the current setup assumes a particular coating, grating, and and quantum efficiency, these are provided to the simulation with a directory path, and can be changed for an engineers specific case. In addition, changing the $N_{ref}$ allows a user to simulate different designs to trade signal-to-noise ratio and total throughput.

Future work will consist of incorporating previously developed models to encompass a wider variety of science. Specifically, Space Telescope Science Institute's \texttt{synphot} simulates spectra based on previous data or developed models{\cite{STScIDevelopmentTeam2018Synphot:Astropy}}. We also plan to incorporate a feature allowing users to calibrate spatial maps to multiple spectra, rather than a single user-input spectrum. This would be useful if a spatial map has multiple sources, and the user has a different input spectrum for each source, for example.

In addition, the current simulation assumes efficiencies and designs of a FUV IFS. SCDDs have called for other modes for the IFS, including NUV and optical. Future iterations of PyISH will include these modes, as well as additional noise sources such as zodiacal background or stray light. Once available, each mode will also include a separate LSF as a user input to incorporate spectrograph broadening.

In conclusion, we believe community use of this tool will achieve the intended goals of exploring the science and engineering trades for including a UV IFS on the HWO instrument suite. Both casual and advanced users alike can utilize this tool to showcase the power of a UV IFS for various science cases. In the future, we hope to incorporate previously developed models such as \texttt{synphot} to simulate a wider variety of science cases that lack defined spectra.

\section{Code, Data, and Materials Availability}
\label{sect: code}

Version (1.0.0) of PyISH is publicly available: \protect{\url{https://github.com/gracesweetak/PyISH}}

\section{Acknowledgments}
\label{sect: Acknowledgements}

I would like Dr. Tracy Becker for her encouragement and time. 

\section{Disclosures}
\label{sect: disc}

The authors declare that there are no financial interests, commercial affiliations, or other potential conflicts of interest that could have influenced the objectivity of this research or the writing of this paper.


\bibliography{references}   
\bibliographystyle{spiejour}   


\vspace{2ex}\noindent\textbf{Grace Sweetak} is a graduate student at Lehigh University, completing her thesis work through a collaboration with NASA Goddard under the advisement of Dr. Breann Sitarski. She received her MS degree in Physics from Lehigh University in 2025, and her BA in physics from Rollins College in 2023. She is on track to graduate with her PhD from Lehigh University in physics in 2028. Her current research interests include UV instrumentation, protostars, and solar system science. She is a member of SPIE.

\vspace{1ex}

\listoffigures
\listoftables

\end{spacing}
\end{document}